\documentclass[12pt]{article}
\usepackage{a4wide,epsfig,amsmath,amssymb,scalefnt}
\usepackage[nosort]{cite}
\usepackage{color}
\usepackage{psfrag}

\newcommand{\hthreel}{{\tt H3m}}

\newcommand{\code}{\tt}

\newcommand{\afourpi}{\frac{\alpha_s}{4\pi}}
\newcommand{\z}[1]{\ensuremath{\zeta({#1})}}
\newcommand{\Msusy}{\ensuremath{m_{\rm SUSY}}}

\newcommand{\epscalar}{$\varepsilon$-scalar}

\newcommand{\abbrev}{\scalefont{.9}}
\newcommand{\drbar}{\ensuremath{\overline{\mbox{\abbrev DR}}}}
\newcommand{\drbarmod}{\ensuremath{\overline{\mbox{\abbrev MDR}}}}

\newcommand{\susy}{{\abbrev SUSY}}

\newcommand{\mtop}{M_t}
\newcommand{\mstop}[1]{m_{\tilde t_{#1}}}

\newcommand{\msquark}{m_{\tilde q}}

\newcommand{\mgluino}{m_{\tilde g}}

\newcommand{\muSUSY}{\mu_{\rm \susy{}}}
\newcommand{\lmmtMS}{l_{tS}}

\newcommand{\lmmtMsq}{l_{t\tilde{q}}}
\newcommand{\lmumt}{l_{{\mu}t}}

\newcommand{\stu}{\tilde{t}_1}
\newcommand{\std}{\tilde{t}_2}
\newcommand{\stl}{\tilde{t}_L}
\newcommand{\str}{\tilde{t}_R}
\newcommand{\Smt}{S_t}
\newcommand{\Cmt}{C_t}

\newcommand{\Mstu}{m_{\tilde{t}_1}}
\newcommand{\Mstd}{m_{\tilde{t}_2}}
\newcommand{\Msti}{m_{\tilde{t}_i}}

\newcommand{\msq}{m_{\tilde{q}}}
\newcommand{\Mes}{M_{\varepsilon}}

\newcommand{\Mgl}{m_{\tilde{g}}}
\newcommand{\Mt}{m_t}
\newcommand{\lMstu}{L_{\mu\tilde{t}_1}}
\newcommand{\lMstd}{L_{\mu\tilde{t}_2}}
\newcommand{\lMsq}{L_{\mu\tilde{q}}}
\newcommand{\lMgl}{L_{\mu\tilde{g}}}
\newcommand{\lMt}{L_{{\mu}t}}
\newcommand{\lMes}{L_{{\mu}\varepsilon}}
\newcommand{\lMsqgl}{L_{\tilde{q}\tilde{g}}}

\newcommand{\as}{\frac{\alpha_s}{\pi}}
\newcommand{\cf}{C_F}
\newcommand{\ca}{C_A}
\newcommand{\Nf}{N_f}
\newcommand{\Nq}{N_q}

\newcommand{\Nt}{N_t}
\newcommand{\TF}{T_F}
\newcommand{\ep}{\epsilon}
\newcommand{\dms}{m_{\tilde{t}_1}^2-m_{\tilde{t}_2}^2}

\newcommand{\abs}[1]{{\ensuremath{\left\lvert {#1} \right\rvert}}}

\newcommand{\Sighat}{\ensuremath{\hat{\Sigma}}}
\newcommand{\sw}{\sin{\vartheta_W}}
\newcommand{\tad}[1]{\ensuremath{t_{\phi_{#1}}}}
\newcommand{\swn}[1]{\sin^{#1}{\vartheta_W}}
\newcommand{\Mst}[1]{\ensuremath{m_{\tilde{t}_{#1}}}}

\begin{document}

\title{\vskip-3cm{\baselineskip14pt
    \begin{flushleft}
      \normalsize SFB/CPP-10-34 \\
      \normalsize TTP/10-23     \\
      \normalsize HU-EP-10/25   \\
      \normalsize WUB/10-13
  \end{flushleft}}
  \vskip1.5cm
  Light MSSM Higgs boson mass to three-loop accuracy
}
\author{\small 
  P. Kant$^{(a)}$,
  R.V. Harlander$^{(b)}$, 
  L. Mihaila$^{(c)}$, 
  M. Steinhauser$^{(c)}$\\[1em]
  {\small\it (a) Institut f\"ur Physik,
    Humboldt-Universit\"at zu Berlin}\\
  {\small\it 
    12489 Berlin, Germany}\\
  {\small\it (b) Fachbereich C, Theoretische Physik,
    Universit{\"a}t Wuppertal}\\
  {\small\it 42097 Wuppertal, Germany}\\
  {\small\it (c) Institut f{\"u}r Theoretische Teilchenphysik, Karlsruhe
    Institute of Technology (KIT)}\\ 
  {\small\it  76128 Karlsruhe, Germany}
}

\date{}

\maketitle

\thispagestyle{empty}

\begin{abstract}

The light CP even Higgs boson mass, $M_h$, is calculated to three-loop
accuracy within the Minimal Supersymmetric Standard Model (MSSM). The
result is expressed in terms of \drbar{} parameters and implemented in
the computer program \hthreel{}.  The calculation is based on the proper
approximations and their combination in various regions of the parameter
space.  The three-loop effects to $M_h$ are typically of the order of a
few hundred MeV and opposite in sign to the two-loop corrections.  The
remaining theory uncertainty due to higher order perturbative
corrections is estimated to be less than 1\,GeV.

 \medskip

\noindent
PACS numbers: 12.60.Jv, 14.80.Da, 12.38.Bx

\end{abstract}

\newpage

\section{\label{sec::intro}Introduction}

Among the main expectations in view of the CERN Large Hadron Collider
(LHC) is to provide clear phenomena beyond the Standard Model (SM) of
particle physics.  A very promising candidate for an extension of the SM
is the so-called Minimal Supersymmetric Standard Model (MSSM) which
relies on an extended symmetry between fermions and
bosons~\cite{Nilles:1983ge,Haber:1984rc}. It is constructed in such
a way that in the low-energy limit the SM is recovered thus leading to
the same phenomena in the energy range around the electroweak scale. In
particular, the MSSM is in accordance with the electroweak precision
data~\cite{Heinemeyer:2004gx}. At the same time it provides a dark
  matter candidate, solves the hierarchy problem and provides a platform
  where also gravitational interactions can be included.

An appealing feature of the MSSM is the quite restrictive Higgs sector which
is described at leading order by two independent parameters. In
particular, the mass 
of the lightest CP-even Higgs boson, $M_h$, is not a free parameter, like in
the SM, but a prediction which can be used in order to test this minimal
supersymmetric extension of the SM.

$M_h$ is very sensitive to radiative corrections. In lowest order it is
bound from above by the $Z$ boson mass which is already excluded by
experiment. Already quite some time ago it has been observed that large
one-loop corrections, in particular from the top quark and top squark
sector can raise $M_h$ to about
140~GeV~\cite{Ellis:1990nz,Okada:1990vk,Haber:1990aw}.  In the meantime
a number of higher order corrections have been computed including even
CP-violating couplings and improvements from renormalization group
considerations (see
Refs.~\cite{Heinemeyer:2004ms,Allanach:2004rh,Frank:2006yh} for a
review).  
In this paper we consider neither CP violation nor the resummation of
higher order logarithms. Let us nevertheless mention that in particular
CP violating phases can lead to a shift of a few GeV in $M_h$, see,
e.g., Refs.~\cite{Heinemeyer:2007aq,Carena:2000yi}.
In Ref.~\cite{Martin:2002wn} a large class of two-loop
corrections to the lightest Higgs boson mass have been considered
and in Ref.~\cite{Martin:2007pg} leading logarithmic corrections at
three-loop order have been computed. The first complete three-loop
calculation of the leading quartic top quark mass terms within
  supersymmetric QCD (SQCD; more precisely, this means supersymmetric
  six-flavor QCD coupled to the MSSM Higgs sector) has been performed in
Ref.~\cite{Harlander:2008ju} for a degenerate supersymmetric mass
spectrum. It is the aim of this paper to provide details and extend this
calculation.

At the moment there are two computer programs publicly available which
include most of the higher order corrections.  {\code FeynHiggs} has
been available already since
1998~\cite{Heinemeyer:1998yj,Degrassi:2002fi,Frank:2006yh} and has been
continuously improved since
then~\cite{Heinemeyer:1998np,Hahn:2009zz}. In particular, it contains
all numerically important two-loop corrections and accepts both real and
complex MSSM input parameters.  The second program, {\code
  CPSuperH}~\cite{Lee:2003nta,Lee:2007gn}, is based on a renormalization
group improved diagrammatic calculation and allows for explicit CP
violation.  Both programs compute the mass spectrum as well as the decay
width of the neutral and charged Higgs bosons.

In this paper we discuss the three-loop corrections originating from the
strong sector of the MSSM which are proportional to the quartic top
quark mass. At three-loop order several mass scales enter the Feynman
diagrams making their evaluation quite involved.  In addition to the top
quark mass there are the top squark masses, the gluino mass and the
masses of the remaining squarks. An exact evaluation of the three-loop
integrals is currently out of range. However, it is possible to apply
expansion techniques for various limits which allow to cover a large
part of the supersymmetric (SUSY) parameter space. In particular, we can
construct precise approximations for the Snowmass Points and Slopes
(SPS)~\cite{Allanach:2002nj,AguilarSaavedra:2005pw}.  Our set-up is
easily extendable to other regions of parameter space which may become
interesting in future.

Together with this paper we provide a {\code Mathematica} program,
\hthreel{}~\cite{h3m}, which contains all our three-loop results. Furthermore,
\hthreel{} constitutes an interface to {\code FeynHiggs}~\cite{FeynHiggs} and various
SUSY spectrum generators which allows for precise predictions of $M_h$ on the
basis of realistic SUSY scenarios.

The remainder of the paper is organized as follows: In the next Section
we revisit the two-loop corrections. In particular, we construct
approximations which are also available at three-loop order and compare
with the exact result.  In Section~\ref{sec::3loops} we provide details
on our three-loop calculations for the various hierarchies. In
particular we discuss the renormalization and the asymptotic expansion.
Section~\ref{sec::h3l} describes the implementation of our results in
the computer program \hthreel{} and the phenomenological implementations
are discussed in Section~\ref{sec::phen}. We present a summary and the
conclusions in Section~\ref{sec::concl}.  In the Appendices additional
material is provided, in particular all the one- and two-loop
counterterms that have entered in our calculation.

\section{\label{sec::2loops}$M_h$ in the MSSM}

\subsection{Higgs boson sector of the MSSM}

The mass of the Higgs boson is obtained from the quadratic terms in the
corresponding potential which for the MSSM has the following form:
\begin{align}
  V_H &= \left( \abs{\muSUSY}^2 + m_1^2 \right) \abs{H_1}^2 
  + \left( \abs{\muSUSY}^2 + m_2^2 \right) \abs{H_2}^2 \notag\\
  &- m_{12}^2 \left(\epsilon_{ab} H_1^a H_2^b + \epsilon_{ab}
    {H_1^a}^*{H_2^b}^*\right) \notag\\ 
  &+ \frac{1}{8} \left(g_1^2 + g_2^2 \right)
  \left[\abs{H_1}^2 - \abs{H_2}^2\right]^2
  + \frac{1}{2} g_2^2 \abs{H_1^\dagger H_2}^2
  \,.
  \label{eq:higgspot}
\end{align}
with $\epsilon_{12}=-\epsilon_{21}=1$ and
$\epsilon_{11}=\epsilon_{22}=0$.  $\mu_{\rm SUSY}$ is the Higgs-Higgsino
bilinear coupling from the super potential and $m_1, m_2$ and $m_{12}$
are soft breaking parameters.  Note that the quartic terms are fixed by
the SU(2) and U(1) gauge couplings $g_1$ and $g_2$.  The parameters in
(\ref{eq:higgspot}) are related to the masses of the gauge bosons and
the pseudoscalar Higgs via
\begin{equation}
  \begin{split}
    M_W^2 &= \frac{1}{2} g_2^2 \left(v_1^2 + v_2^2\right), \\
    M_Z^2 &= \frac{1}{2}\left( g_1^2 + g_2^2 \right)\left( v_1^2 + v_2^2
    \right),\\
    M_A^2 &= - m_{12}^2\left( \tan\beta + \cot\beta \right)\,,
  \end{split}
\end{equation}
where $\tan\beta=v_2/v_1$.

After spontaneous symmetry breaking the neutral components of the
doublets $H_1$ and $H_2$ acquire the vacuum expectation values $v_1$ and
$v_2$, and we write
\begin{align}
  \label{eq:H12phi12}
  H_1 &=
  \begin{pmatrix}
    v_1 + \frac{1}{\sqrt{2}} \left(\phi_1 + i \chi_1 \right) \\
    - \phi_1^-
  \end{pmatrix}\,,\notag\\
  H_2 &=
  \begin{pmatrix}
    \phi_2^+ \\
    v_2 + \frac{1}{\sqrt{2}} \left( \phi_2 + i \chi_2 \right)
  \end{pmatrix}\,,
\end{align}
which leads to the following representation of the Higgs boson mass matrix
at tree level:
\begin{eqnarray}
  {\cal M}_{H,\rm tree}^2 &\!=\!&
  \frac{1}{2} \frac{\partial^2 V}{\partial \phi_1 \partial \phi_2}
  =
  \frac{\sin 2\beta}{2}
  \left(
  \begin{array}{cc}
    M_Z^2 \cot\beta + M_A^2 \tan\beta &
    -M_Z^2-M_A^2 \\
    -M_Z^2-M_A^2 &
    M_Z^2 \tan\beta + M_A^2 \cot\beta
  \end{array}
  \right)
  \,,
  \label{eq::MHtree}
\end{eqnarray}
where we restrict ourselves to the two CP-even Higgs bosons.

In this paper we restrict ourselves to the numerically dominant
$m_t^4$ corrections. As a consequence we can set the electroweak gauge
couplings to zero and furthermore nullify the external momentum in the
occurring two-point functions. The formalism presented in the following
is adapted to this framework.

\begin{figure}[t]
  \centering
  \begin{tabular}{cc}
    \includegraphics[width=.45\linewidth]{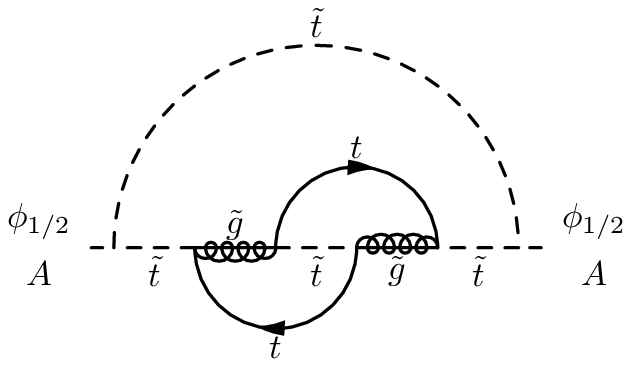}
    &\quad
    \raisebox{-.1em}{\includegraphics[width=.45\linewidth]{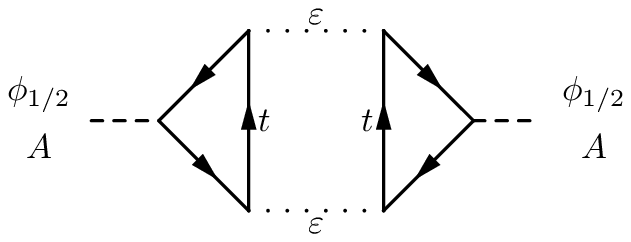}}
    \\
    (a) & (b)
    \\[1em]
    \includegraphics[width=.45\linewidth]{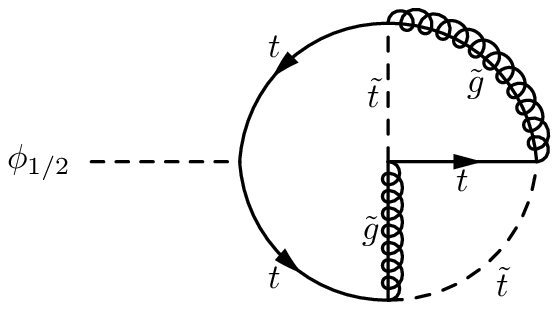}
    &\quad
    \raisebox{-.7em}{\includegraphics[width=.45\linewidth]{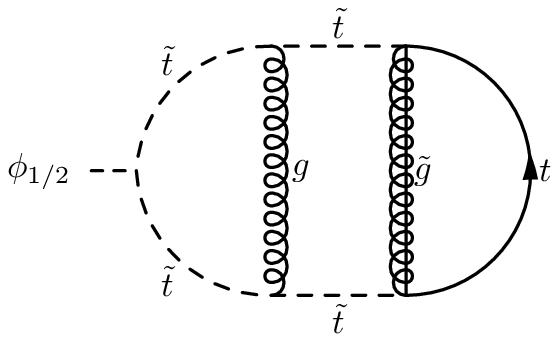}}
    \\
    (c) & (d)
  \end{tabular}
  \caption{Sample diagrams contributing to $\Sigma_{\phi_1}$,
    $\Sigma_{\phi_2}$, $\Sigma_{\phi_1\phi_2}$, 
    $\Sigma_A$, $t_{\phi_1}$ and $t_{\phi_2}$.
    Internal solid, dashed, dotted and curly lines correspond to
    top quarks, top squarks, \epscalar{}
    and gluons, respectively. Gluinos are
    depicted with as curly lines with an additional solid line in the middle.
    The external dashed line corresponds to the Higgs boson.
  }
  \label{fig::diags}
\end{figure}

It is convenient to evaluate the quantum corrections in the
$\{\phi_1,\phi_2\}$ basis which requires the evaluation of self energy
corrections $\Sigma_{\phi_1}$ and $\Sigma_{\phi_2}$ involving $\phi_1$
and $\phi_2$. Denoting renormalized quantities with a hat,
Eq.~(\ref{eq::MHtree}) gets modified to
\begin{eqnarray}
  {\cal M}_{H}^2 &=&
  {\cal M}_{H,\rm tree}^2 -
  \left(
  \begin{array}{cc}
    \hat\Sigma_{\phi_1}       & \hat\Sigma_{\phi_1\phi_2} \\
    \hat\Sigma_{\phi_1\phi_2} & \hat\Sigma_{\phi_2}
  \end{array}
  \right)
  \,,
  \label{eq::MH}
\end{eqnarray}
with\cite{Heinemeyer:1998np}
\begin{align}
  \Sighat_{\phi_1} &= \Sigma_{\phi_1} \nonumber
  \begin{aligned}[t]
    &- \Sigma_A \sin^2\beta\\
    &+ \frac{e}{2M_W\sw} \tad1 \cos\beta\left( 1 + \sin^2\beta \right)\\
    &- \frac{e}{2M_W\sw} \tad2 \cos^2\beta \sin\beta \,,
  \end{aligned}\\
  \Sighat_{\phi_2} &= \Sigma_{\phi_2} \nonumber
  \begin{aligned}[t]
    &- \Sigma_A \cos^2\beta\\
    &- \frac{e}{2M_W\sw} \tad1 \sin^2\beta \cos\beta\\
    &+ \frac{e}{2M_W\sw} \tad2 \sin\beta \left( 1 + \cos^2\beta \right) \,,
  \end{aligned}\\
  \Sighat_{\phi_1\phi_2} &= \Sigma_{\phi_1\phi_2}
  \begin{aligned}[t]
    &+ \Sigma_A \sin\beta \cos\beta\\
    &+ \frac{e}{2M_W\sw} \tad1 \sin^3\beta\\
    &+ \frac{e}{2M_W\sw} \tad2 \cos^3\beta \,.
  \end{aligned}
\end{align}
In this equation, $\vartheta_W$ is the weak mixing angle, $\Sigma_A$
denotes the self energy of the pseudo-scalar Higgs boson and
$t_{\phi_i}$ the tadpole contributions of the field $\phi_i$.  Typical
diagrams to the individual contributions can be found in
Fig.~\ref{fig::diags}.  Since we are only interested in the leading
corrections proportional to $m_t^4$ we evaluate the quantum corrections
in the limit of vanishing external
momentum~\cite{Chankowski:1991md,Brignole:1992uf,Dabelstein:1994hb}.

For the evaluation of the lightest Higgs boson mass we consider in a
first step the matrix element of ${\cal M}_{H}^2$ to a given order in
perturbation theory. Subsequently, we determine the eigenvalues and
assign the smaller one to $M_h$. We perform this procedure at tree level
and at one-, two-, and three-loop order which leads to the corresponding
approximations of the Higgs boson mass. It is convenient to introduce
the quantity
\begin{eqnarray}
  \Delta M_h^{(i)} &=& M_h^{(i-{\rm loop})} - M_h^{\rm tree}
  \,,
\end{eqnarray}
representing the difference between the Higgs boson mass evaluated with
$i$-loop accuracy and the tree-level result.

\subsection{Top squark sector of the MSSM}

In this paper we only consider strong corrections which means that apart from
the quarks and gluons also the corresponding superpartners, the squarks and
gluinos, are present. The leading contribution proportional to $G_F m_t^4$
is generated by the top quark Yukawa coupling which distinguishes
the top squark sector from the other squark parts of the MSSM Lagrange
density. In order to fix the notation let us discuss in more detail the mass
matrix of the left- and right-handed component of the top squark, $\tilde t_L$
and $\tilde t_R$, which is given by
\begin{align}
  {\cal M}_{\tilde{t}}^2 &= 
  \begin{pmatrix}
    \Mt^2 
    + M_Z^2 \left(
      \tfrac{1}{2}-\tfrac{2}{3} \swn{2} \right) 
    \cos 2\beta
    + M_{\tilde{Q}}^2
    &  \Mt \left(A_t - \muSUSY\cot\beta\right) 
    \\ \Mt \left(A_t - \muSUSY\cot\beta\right) 
    &  \Mt^2 
    + \tfrac{2}{3} M_Z^2 \swn{2}\cos 2\beta
    + M_{\tilde{U}}^2
  \end{pmatrix}
  \notag\\&\equiv
  \begin{pmatrix}
    \Mst{L}^2
    & \Mt X_t
    \\ \Mt X_t
    & \Mst{R}^2
  \end{pmatrix}\,,
  \label{eq::mstop}
\end{align}
with $X_t=A_t-\mu_{\rm SUSY} \cot\beta$.
$M_{\tilde Q}$ and $M_{\tilde U}$ are soft SUSY breaking masses, and $A_t$ is the
soft SUSY breaking tri-linear coupling between the Higgs boson and the top
squark fields.

Diagonalization of Eq.~(\ref{eq::mstop}) leads to the mass eigenstates
$\tilde{t}_1$ and $\tilde{t}_2$ with masses
\begin{equation}
  \Mst{1,2}^2 = \frac{1}{2}\left(
    \Mst{L}^2 + \Mst{R}^2
    \mp \sqrt{
      \left( \Mst{L}^2 - \Mst{R}^2 \right)^2
      + 4 \Mt^2 X_t^2
    }
  \right)\,.
  \label{eq:mst12}
\end{equation}

The mixing angle is defined through the unitary transformation
\begin{eqnarray}
\left(
\begin{array}{cc}
m_{\tilde{t}_1}^{2}& 0\\
0& m_{\tilde{t}_2}^{2}
\end{array}
\right) = {\cal R}_{{\tilde t}}^{\dag} {\cal M}_{{\tilde t}}{\cal
  R}_{{\tilde t}}\,,\quad \mbox{with} \quad{\cal R}_{{\tilde t}} =\left(
\begin{array}{cc}
\cos \theta_{t} & -\sin \theta_{t} \\
\sin \theta_{t} & \cos \theta_{t}
\end{array}
\right)\,,
\label{eq::mixa}
\end{eqnarray}
and 
\begin{eqnarray}
\sin 2\theta_{t} =  \frac{2 m_t (A_t-\mu_{\rm
    SUSY}\cot\beta)}{m_{\tilde{t}_1}^2-m_{\tilde{t}_2}^2} \,.
\label{eq::Atmu}
\end{eqnarray}

\subsection{Leading $m_t^4$ corrections in the on-shell and $\overline{\rm
    DR}$ scheme}

As already mentioned above, the numerically dominant contribution arises
from the self energy diagrams evaluated for vanishing external
momentum. Thus, it is convenient to introduce the following notation for
the $i$-loop corrections to the Higgs boson mass
\begin{eqnarray}
  \Delta M_h^{(i)} &=& \Delta^{m_t^4} M_h^{(i)} + \Delta^{\rm rem} M_h^{(i)}
  \,,
\end{eqnarray}
where $\Delta^{m_t^4} M_h^{(i)}$ comprises the complete SQCD
contribution of order $\alpha_t\alpha_s^{i-1}$ ($\alpha_t$ is the top
Yukawa coupling) originating from the top quark/squark sector for
vanishing external momentum which is proportional to $m_t^4$.  At
  one-loop order only top quarks and top squarks are present in the
  loops. The two- and three-loop corrections, $\Delta^{m_t^4}
  M_h^{(2)}$ and $\Delta^{m_t^4} M_h^{(3)}$, are obtained by adding
  gluon, gluino, quark and squark contributions.  $\Delta^{\rm rem}
M_h^{(i)}$ represents the remaining part which is only available at one-
and two-loop order.  In our approach these corrections are taken from
{\code FeynHiggs} which includes the complete one-loop corrections and
all available two-loop terms.\footnote{For a detailed description we
  refer to the {\code FeynHiggs} home page~\cite{FeynHiggs}.}  At
three-loop order only the contribution $\Delta^{m_t^4} M_h^{(3)}$ is
considered.

In the following we discuss the relative contribution to
the Higgs boson mass comparing $\Delta M_h^{(i)}$ and $\Delta^{m_t^4} M_h^{(i)}$
at one- and two-loop order ($i=1,2$) where
both the on-shell and \drbar{} scheme for the mass parameters and
the mixing angle are considered.
For illustration we adopt the scenarios SPS1a and SPS2 and show the Higgs
boson mass as a function of $m_{1/2}$.

\begin{figure}
  \centering
  \begin{tabular}{cc}
    \includegraphics[width=.45\textwidth]{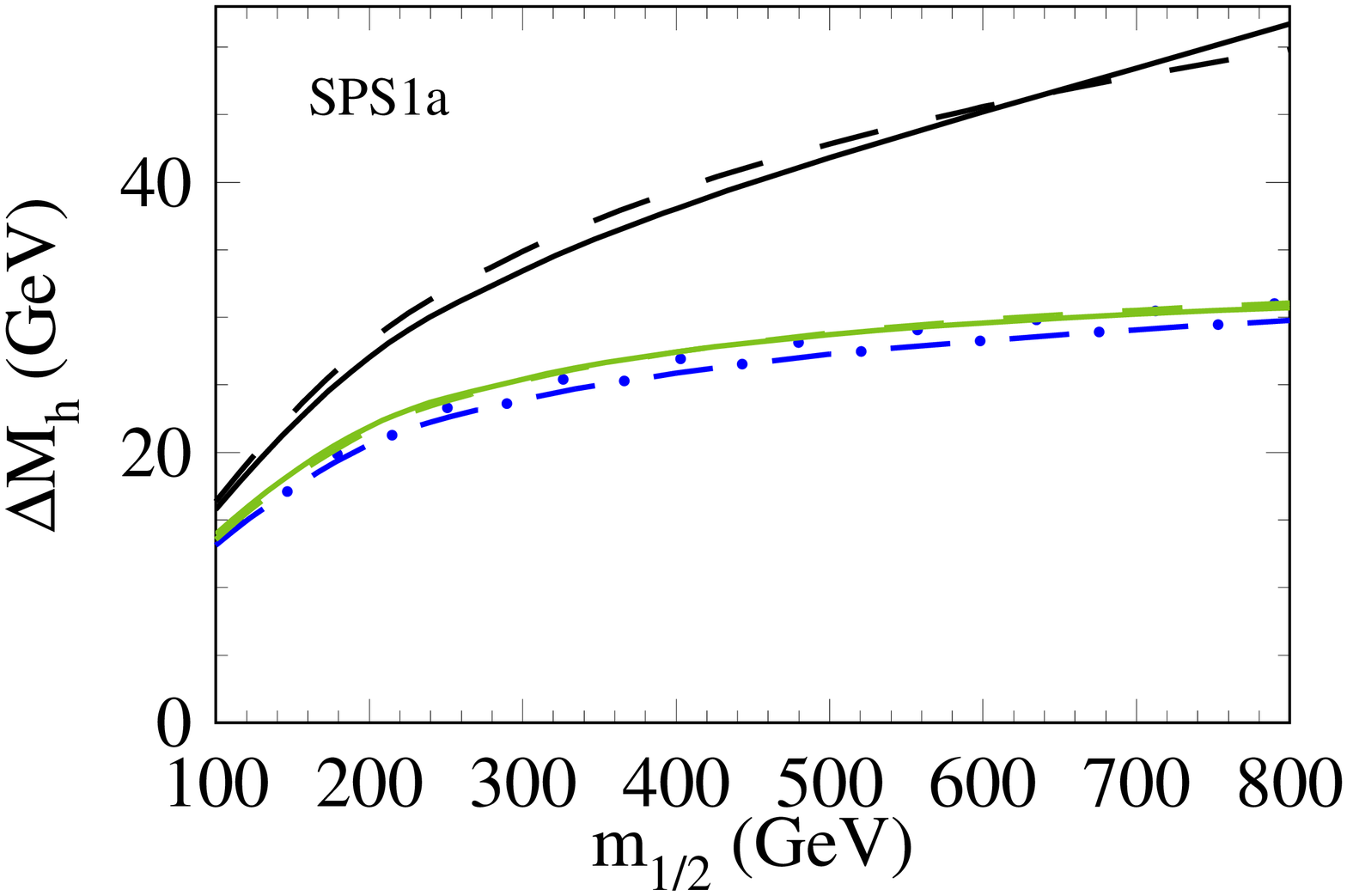}
    &
    \includegraphics[width=.45\textwidth]{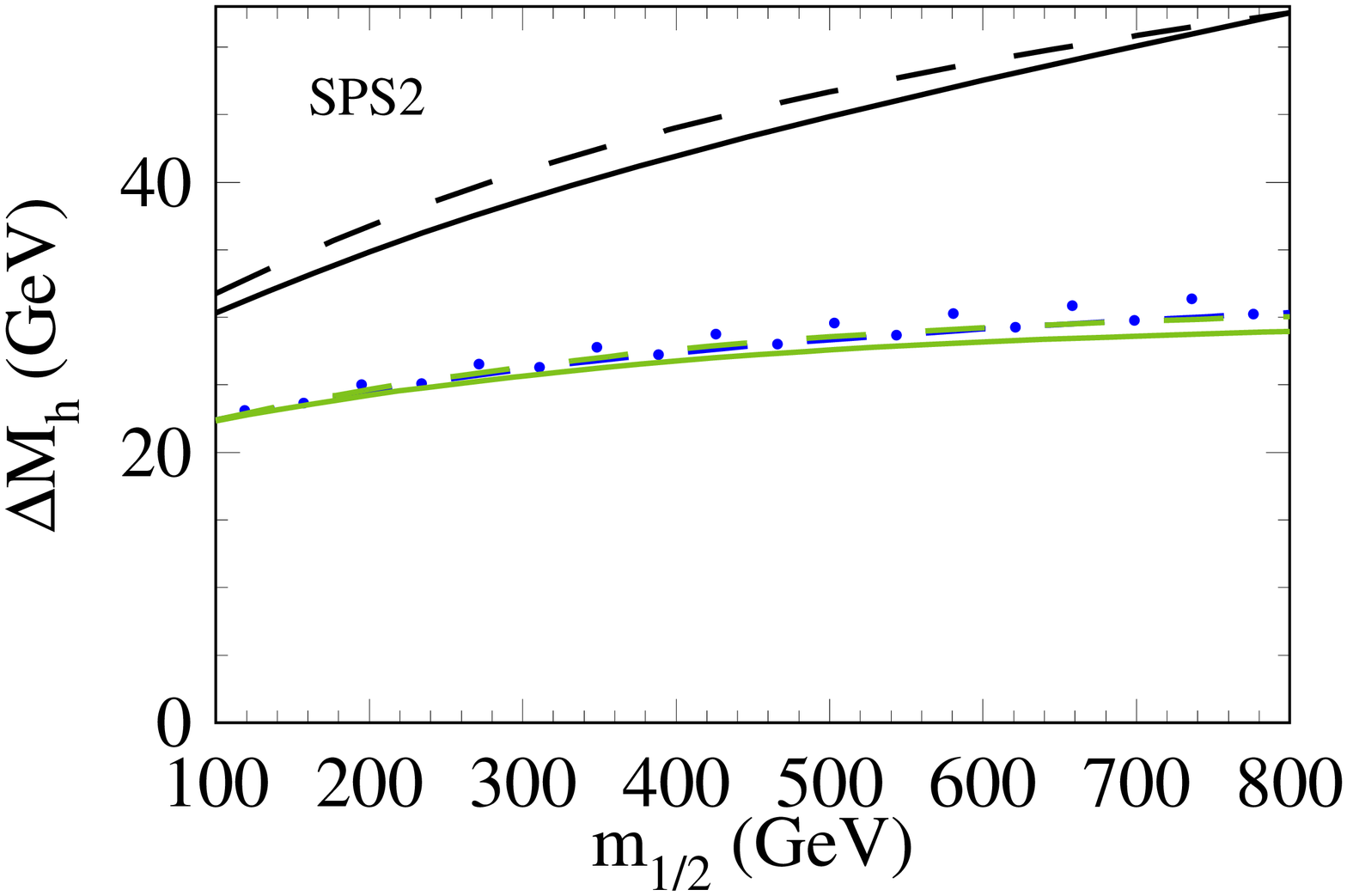}
    \\ (a)&(b)\\
  \end{tabular}
  \caption{\label{fig::mt4}Comparison of complete and approximate one-
    and two-loop corrections to the Higgs boson mass for SPS1a (a) and
    SPS2 (b).  The solid (full result) and dashed lines ($m_t^4$
    approximation) represent the results in the on-shell scheme where
    the upper and lower curves correspond to the one- and two-loop
    results, respectively.  The two-loop \drbar{} results are shown as
    dash-dotted (full result) and dotted ($m_t^4$ approximation) curves.
  }
\end{figure}

The solid lines in Fig.~\ref{fig::mt4} show the results for $\Delta
M_h^{(i)}$ using on-shell parameters\footnote{For the on-shell
  renormalization of the mixing angle we adopt the convention of
  Ref.~\cite{Degrassi:2001yf}.} for the masses and $\theta_t$ as it is
provided by {\code FeynHiggs}. The dashed lines correspond to the
$m_t^4$ approximation ($\Delta^{m_t^4} M_h^{(i)}$) which can be found in
Refs.~\cite{Heinemeyer:1998np,Degrassi:2001yf} and has been confirmed by
us by an independent calculation. The panels (a) and~(b) correspond to
the SPS1a and SPS2 benchmark scenarios, respectively, where the input
parameters have been generated with the help of {\code
  SOFTSUSY}~\cite{Allanach:2001kg}.  The small differences of the solid
and dashed lines\footnote{In the case of SPS1a the two-loop dashed line is
  almost on top of the solid one.} demonstrate that the leading top
quark mass term approximates the full result to a high accuracy.  This
statement is also true in the \drbar{} scheme. The corresponding
two-loop results are shown in Fig.~\ref{fig::mt4} as dash-dotted and
dotted curves where the former corresponds to the full and the latter to
the approximate result.

It is well known that the perturbative series can exhibit a bad
convergence behaviour in case it is parametrized in terms of the
on-shell quark masses\footnote{For a typical example we refer to the
  electroweak $\rho$ parameter. Using the on-shell top quark mass the
  four-loop
  corrections~\cite{Schroder:2005db,Chetyrkin:2006bj,Boughezal:2006xk}
  are larger by a factor 50 as compared to the $\overline{\rm MS}$
  scheme.}  which is due to intrinsically large contributions related to
the infra-red behaviour of the theory.  Thus, it is tempting to
re-parametrize the results for the Higgs boson mass in terms of
\drbar{} parameters for the top quark mass, the masses of the
SUSY particles and the top squark mixing angle.  Since the corrections
are dominated by $\Delta^{m_t^4} M_h^{(i)}$ it is sufficient to consider
only this term and take over $\Delta^{\rm rem} M_h^{(i)}$ from the
output of {\code FeynHiggs}.  In the remainder of the paper we will
refer to this renormalization scheme as \drbar{} scheme although it
contains a mixture of on-shell and \drbar{} parameters.  Let us mention
already at this point that from outside this mixture does not pose any
complication in the practical use since the spectrum generator produces
both on-shell and \drbar{} parameters which then serve as input for the
evaluation of $M_h$.

Further below we will discuss the scheme dependence of $M_h$ and indeed show
that the loop corrections are in general smaller in the \drbar{}
scheme (cf. Fig.~\ref{fig::DRvsOS}). Similar studies can also be found in the
literature~\cite{Degrassi:2002fi,Allanach:2004rh}.

The considerations of this subsection motivates the following procedure
at three loops: It is certainly sufficient to consider only the
approximation $\Delta^{m_t^4} M_h^{(3)}$ since the size of the remaining
term is expected to be below 100~MeV.  Furthermore, we adopt the
\drbar{} scheme since we expect that the perturbative series
shows a better convergence behaviour. In addition the evaluation of the
counterterms themselves is significantly simpler. Actually, most of them
are already available in the literature and the computation of the
remaining ones is quite straightforward as we discuss in Appendix~A. We
provide the analytical results for the two-loop renormalization
constants of the top squark masses and mixing angle, that can be easily
expanded for the mass hierarchies considered in this paper. The
  multiplicative \drbar{} renormalization constants of the
top quark and gluino mass are mass independent and therefore valid for
all hierarchies.

\subsection{Construction of approximations}

Considering the many different mass parameters entering the formula for
the Higgs boson mass an exact calculation of the three-loop corrections
is currently not feasible. However, due to the various hierarchies among
the particle masses it is promising to consider expansions in properly
chosen small parameters. As a guideline for the latter we follow the SPS
scenarios as defined in
Refs.~\cite{Allanach:2002nj,AguilarSaavedra:2005pw}.

In order to construct approximations covering all SPS cases it is
sufficient to consider the following hierarchies among the SUSY
masses\footnote{We decided to keep the nomenclature for the hierarchies
  as they are in our internal computations and documents.  The
  non-continuous numeration results from the fact that for testing
  purposes we have computed further hierachies which, however, are not
  included in the program \hthreel{} (cf. Section~\ref{sec::h3l}).}
\begin{eqnarray}
  ({\rm h3}) && m_{\tilde q}\approx m_{\tilde t_1}\approx m_{\tilde t_2}\approx m_{\tilde g}
  \,,\nonumber\\
  ({\rm h4}) && m_{\tilde q} \gg m_{\tilde t_1} \approx m_{\tilde t_2} \approx m_{\tilde g}
  \,,\nonumber\\
  ({\rm h5}) && m_{\tilde q} \gg m_{\tilde t_2} \gg m_{\tilde t_1} \approx m_{\tilde g}
  \,,\nonumber\\
  ({\rm h6}) && m_{\tilde q} \gg m_{\tilde t_2} \approx m_{\tilde g} \gg m_{\tilde t_1} 
  \,,\nonumber\\
  ({\rm h6b}) && m_{\tilde q} \approx m_{\tilde t_2} \approx m_{\tilde g}\gg m_{\tilde t_1} 
  \,,\nonumber\\
  ({\rm h9}) && m_{\tilde q} \approx m_{\tilde t_1} \approx m_{\tilde t_2} \gg m_{\tilde g} 
  \,,
  \label{eq::hierarchies}
\end{eqnarray} 
where in the case of ``$\gg$'' an asymptotic expansion in the
corresponding hierarchy is performed. In the case of ``$\approx$'' a
naive Taylor expansion in the difference of the particle masses is
sufficient.  Throughout this paper, $\tilde q$ denotes any squark other
than $\tilde t$, and we assume a common mass value $m_{\tilde q} \equiv
m_{\tilde{q}_1} = m_{\tilde{q}_2}$ for all of these ``heavy squarks''.

In all hierarchies we assume that the SUSY masses are larger than
the top quark mass and perform an asymptotic expansion in the corresponding
ratio.
In the numerical results discussed below we include for the various
hierarchies the expansion terms as given in Tab.~\ref{tab::exp}
where the following notation has been introduced
\begin{eqnarray}
  x_{12} &=& \frac{m_{\tilde t_1}}{m_{\tilde t_2}}
  \,,\nonumber\\
  x_{2g} &=& \frac{m_{\tilde t_2}}{m_{\tilde g}}
  \,,\nonumber\\
  x_{1g} &=& x_{12} x_{2g} = \frac{m_{\tilde t_1}}{m_{\tilde g}}
  \,,\nonumber\\
  x_{1q}&=& \frac{m_{\tilde t_1}}{m_{\tilde q}}
  \,,\nonumber\\
  x_{2q}&=& \frac{m_{\tilde t_2}}{m_{\tilde q}}
  \,.
\end{eqnarray}

\begin{table}
  \begin{center}
    \begin{tabular}{l|lll}
      hierarchy & \multicolumn{3}{l}{expansion depth}
      \\
      \hline
      (h3) & $(1-x_{12}^2)^3$, & $(1-x_{1g})^3$, & $(1-x_{1q}^2)^3$ 
      \\
      (h4) & & & $x_{1q}^8$ 
      \\
      (h5) & $(1-x_{1g})^2$, & $(x_{12})^4$, & $x_{2q}^4$ 
      \\
      (h6) & $x_{12}^3$, & $(1-x_{2g})^2$, & $x_{2q}^4$ 
      \\
      (h6b) &$x_{12}^3$, & $(1-x_{2g})^2$, & $(1-x_{2q}^2)^2$
      \\
      (h9) & $(1-x_{12}^2)^3$, & $1/x_{12}^4$, & $(1-x_{1q}^2)^3$ 
    \end{tabular}
    \caption{\label{tab::exp}Expansion terms available for the individual
      hierarchies as defined in Eq.~(\ref{eq::hierarchies}) at three-loop order.}
  \end{center}
\end{table}

Note that at two-loop order the contributions involving the squarks
$\tilde q$ with $q\in\{u,d,s,c,b\}$ cancel in the sum of all
diagrams. At three-loop level, however, the results depend on $m_{\tilde
  q}$.  In those cases where $m_{\tilde q}$ is much larger than the
other masses at least three expansion terms are computed and a good
convergence even up to $m_{\tilde q} \approx m_{\tilde t_2}$ is
observed.

\begin{figure}[t]
  \centering
  \begin{tabular}{c}
    \includegraphics[width=\textwidth]{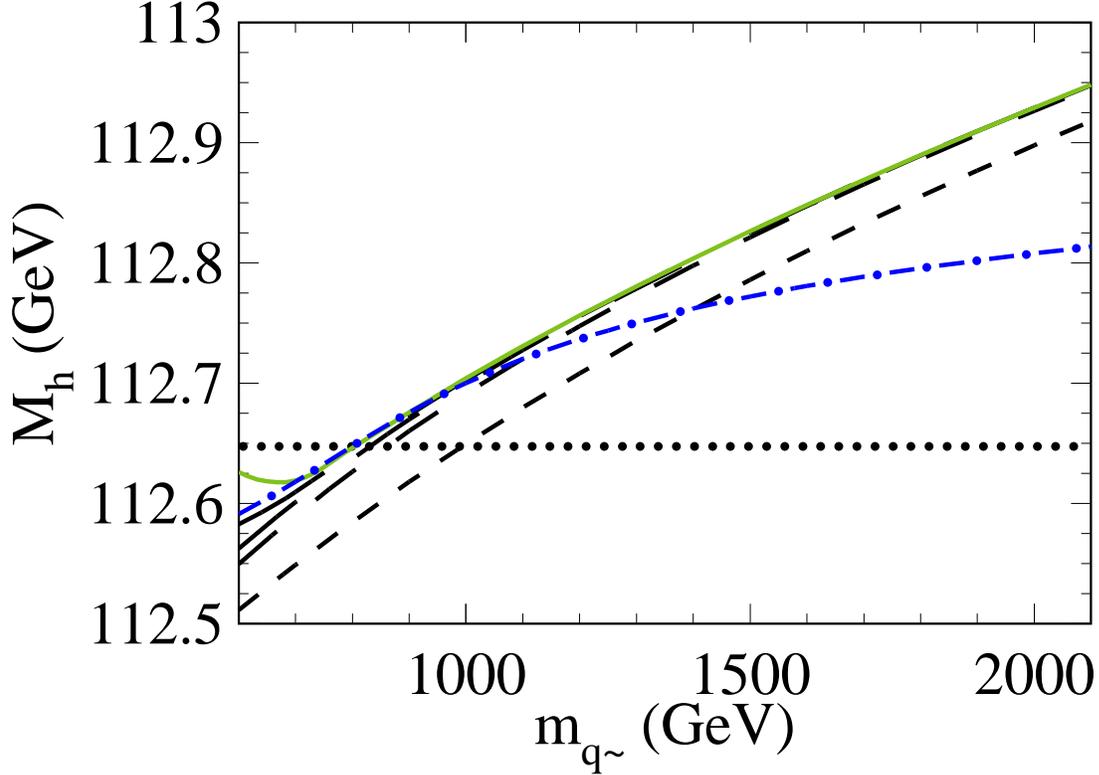}
  \end{tabular}
  \caption{\label{fig::large_mgtilde}Dependence of the three-loop corrections
    on the heavy squark mass $m_{\tilde{q}}$. 
    The dashed and solid lines correspond to the
    hierarchy (h4) where successively higher order terms in 
    $m_{\rm SUSY}/m_{\tilde{q}}$ have been included (The solid curve contains
    terms of order $(m_{\rm SUSY}^2/m_{\tilde{q}}^2)^{4}$.). The dotted and
    dash-dotted curves correspond to the hierarchy (h3); for the dotted
    line $m_{\tilde{q}}$ has been kept fixed at 800~GeV, the dash-dotted curve
    includes terms of order $(1-m_{\rm SUSY}^2/m_{\tilde{q}}^2)^{3}$.
    }
\end{figure}

In order to demonstrate this point we consider the hierarchies (h3) and
(h4) and show in Fig.~\ref{fig::large_mgtilde} the three-loop prediction
for $M_h$. The dashed and solid lines correspond to (h4) including
successively higher orders in $1/m_{\tilde{q}}$ where for illustration
the following input parameters have been chosen:\footnote{If not stated
  otherwise we set the renormalization scale equal to the on-shell top
  quark mass and evaluate all \drbar{} parameters at that
  scale. Note, however, that \hthreel{} is not restricted to this
  choice.}
\begin{eqnarray}
  m_{\rm SUSY} &\equiv& m_{\tilde{g}} 
  \,\,=\,\, m_{\tilde{t}_1} \,\,=\,\, m_{\tilde{t}_2} \,\,=\,\, 800\,\text{GeV}
  \,,\nonumber\\
  A_t &=& \mu_{\rm SUSY} \,\,=\,\, \theta_t \,\,=\,\, 0
  \,,\nonumber\\
  M_A &=& 1500~\mbox{GeV}\,.
\end{eqnarray}
The horizontal ($m_{\tilde{q}}$-independent) dotted line corresponds to
the scenario (h3) with $m_{\tilde{q}}=m_{\rm SUSY}$ fixed at 800~GeV.
One observes a crossing of the latter and the (h4)-curve including
$1/m_{\tilde{q}}^8$ corrections for $m_{\tilde{q}}\approx m_{\rm SUSY}$
which nicely demonstrates the rapid convergence in the
large-$m_{\tilde{q}}$ expansion.  Fig.~\ref{fig::large_mgtilde} also
shows that the expansion around $m_{\tilde{q}}=m_{\rm SUSY}$ leads to
good approximations even if $m_{\tilde{q}}$ is two to three times as big
as $m_{\rm SUSY}$, see dash-dotted curve.  Note that for this plot we
computed the \drbar{} top quark mass for $\msq=800$~GeV and kept it
fixed.

Let us in a next step compare the approximate SQCD corrections according
to our hierarchies with the full prediction for $M_h$ from
Ref.~\cite{Degrassi:2001yf}.  For illustration we adopt in the remainder
of this Section a minimal supergravity ({\sc msugra}) scenario with
\begin{eqnarray}
  \tan\beta &=&10\,,\nonumber\\
  A_0&=&0\,,\nonumber\\
  \mu_{\rm SUSY}&>&0\,,
  \label{eq::tanbeta}
\end{eqnarray}
and vary $m_0$ and $m_{1/2}$ as follows
\begin{eqnarray}
  60~\mbox{GeV}  & < m_0     < & 1600~\mbox{GeV}\,,\nonumber\\
  100~\mbox{GeV} & < m_{1/2} < & 800~\mbox{GeV}\,.
  \label{eq::m0m12range}
\end{eqnarray}
We have checked that very similar results are obtained 
for other choices of $\tan\beta, A_0$, and sign$(\mu_{\rm SUSY})$. Thus, our conclusions
are at least valid for all {\sc msugra} SPS scenarios (cf. Appendix~C). 

\begin{figure}[t]
  \centering
  \begin{tabular}{cc}
    \includegraphics[width=.45\textwidth]{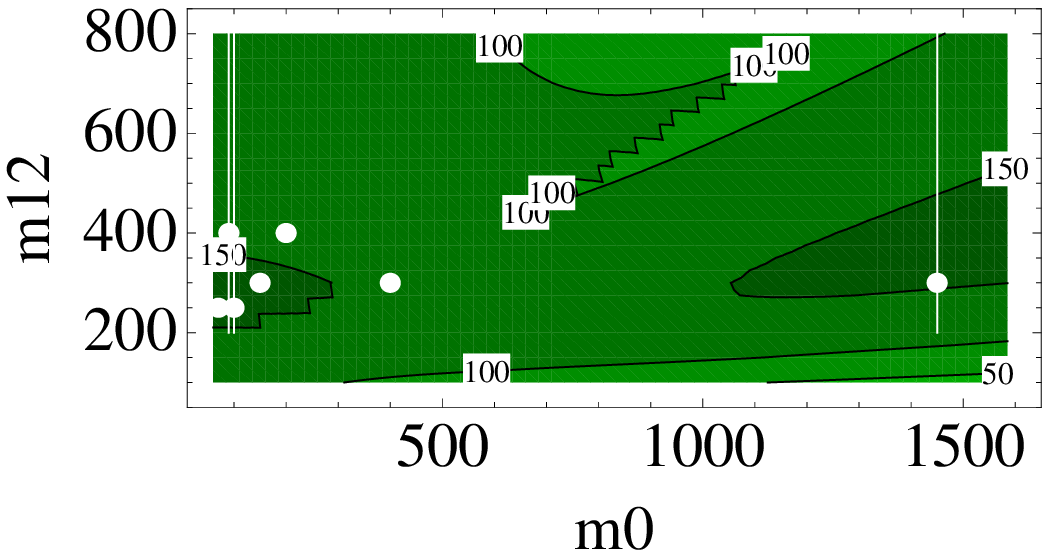}
    &
    \includegraphics[width=.45\textwidth]{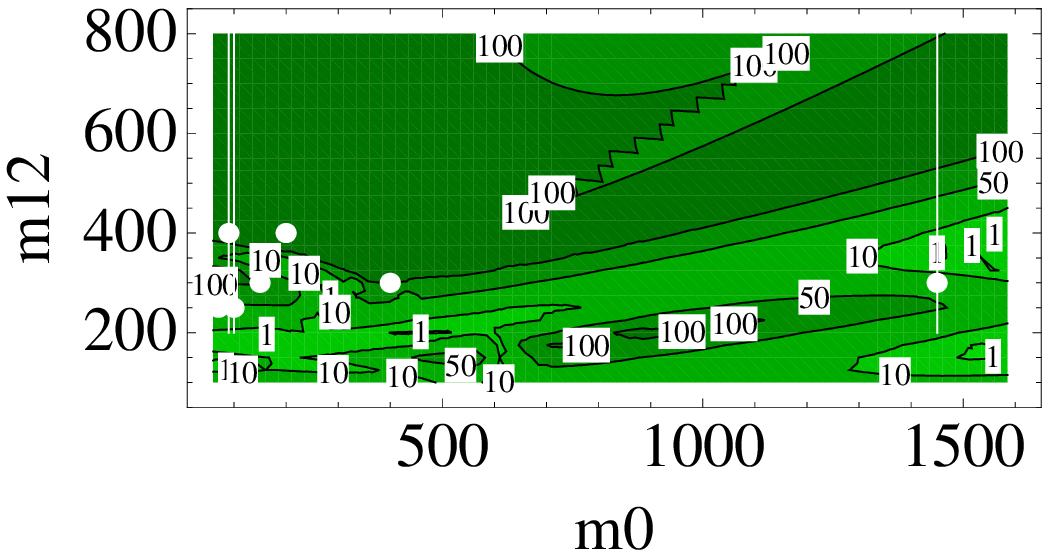}
    \\
    (a) & (b) \\
    \includegraphics[width=.45\textwidth]{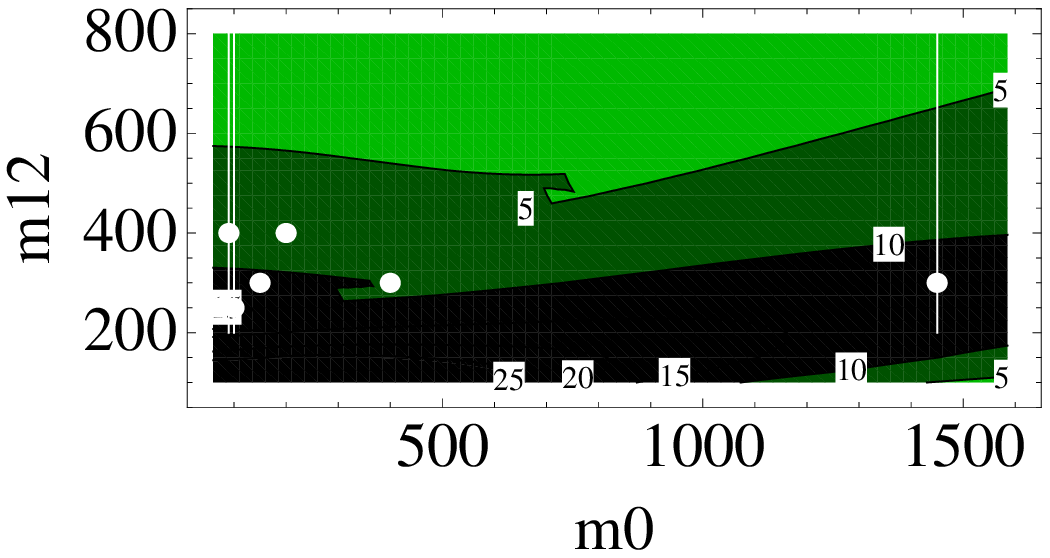}
    &
    \includegraphics[width=.45\textwidth]{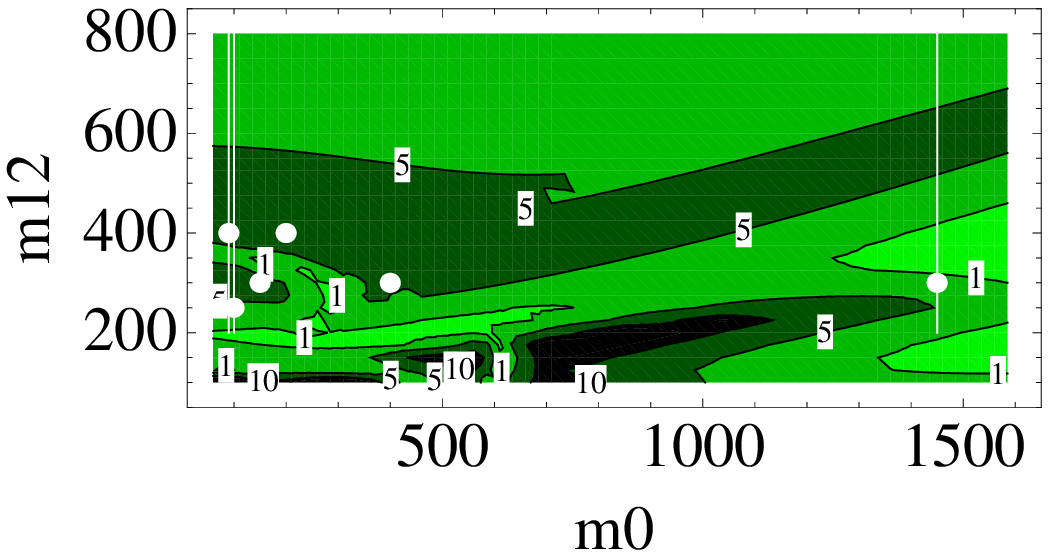}
    \\
    (c) & (d)
  \end{tabular}
  \caption{\label{fig::2lappr} Comparison of approximate and full
    two-loop result for hierarchy (h3) (a) and the combination of (h3),
    (h5), (h6), (h6b) and (h9) (b). The contour lines indicate the
    deviations in MeV.  In (c) and (d) the results of (a) and (b) are
    normalized to the genuine two-loop contributions where the contour
    lines indicate the deviations in per cent. The benchmark points and
    slopes are shown as (white) dots and lines.  }
\end{figure}

In Fig.~\ref{fig::2lappr}(a) and~(b) the absolute value of the
difference between the full and the approximate two-loop prediction for
the Higgs boson mass, $M_h^{(2)} - M_h^{(2),\rm 3lcut}$, is shown in the
$m_0$-$m_{1/2}$ plane where $M_h^{(2),\rm {3lcut}}$ includes the same
number of expansion terms which are available at three loops.  In
Fig.~\ref{fig::2lappr}(a) we only include the results from hierarchy
(h3) whereas in (b) also (h5), (h6), (h6b) and (h9) enter. {For each
  hierarchy we compute the difference to the exact result and plot in
  Fig.~\ref{fig::2lappr}(b) the minimum. We define the relative
  uncertainty through}
\begin{eqnarray}
  \delta^{(2)} &=& \frac{M_h^{(2)} - M_h^{(2),\rm 3lcut}}{M_h^{(2)}-M_h^{(1)}}
  \,,
  \label{eq::delta2}
\end{eqnarray}
which is shown in Fig.~\ref{fig::2lappr}(c) and (d). In
Eq.~(\ref{eq::delta2}) $M_h^{(i)}$ corresponds to the exact $i$-loop
prediction.  For reference we show in Fig.~\ref{fig::2lappr} the {\sc
  msugra} SPS benchmark points and slopes as (white) dots and lines,
having in mind that for some of them the values of $\tan\beta$ and $A_0$
are different from the ones chosen in Eq.~(\ref{eq::tanbeta}).  The
assignment of the individual scenarios to the corresponding dot is
easily done with the help of the table in Appendix~C.

Already for (h3) alone one observes a good coverage in the whole $m_0$-$m_{1/2}$
plane with deviations smaller than 150~MeV.
This gets further improved after including the other hierarchies.
For lower values of $m_{1/2}$ one has relative deviations also above 10\%,
however, the absolute difference between the full result and the
approximation is below 100~MeV.

In Tab.~\ref{tab::SPS} we directly compare the two-loop predictions for $M_h$
for the benchmark points listed in the table in Appendix~C 
and SPS7 and SPS8
(gauge-mediated supersymmetry breaking)~\cite{Allanach:2002nj}.
As before, the full results are based on {\code FeynHiggs} and
Ref.~\cite{Degrassi:2001yf}, and for $M_h^{(2),\rm appr}$ we use the
approximation incorporated in \hthreel{}.
An impressive agreement is found, often even below 100~MeV.

\begin{table}[t]
  \begin{center}
    \begin{tabular}{c|l|l|c}
      & $M_h^{(2)}$ & $M_h^{(2),\rm appr}$ & optimal \\
      & (GeV)       & (GeV)               & hierarchy \\
      \hline
SPS1a &  111.81 &  111.84 & h6b\\
SPS1a$^\prime$ &  113.26 &  113.27 & h6b\\
SPS1b &  115.53 &  115.64 & h3\\
SPS2 &  115.65 &  115.77 & h5\\
SPS3 &  114.63 &  114.77 & h3\\
SPS4 &  113.73 &  113.77 & h6\\
SPS5 &  111.66 &  111.83 & h3\\
SPS7 &  112.20 &  112.21 & h3\\
SPS8 &  114.19 &  114.20 & h3\\
    \end{tabular}
    \caption{\label{tab::SPS}Comparison of full and approximate two-loop
      prediction for $M_h$ for the different benchmark points.}
  \end{center}
\end{table}

The results discussed in this Subsection are very promising in view of
the three-loop approximation. At two-loop order the expansion terms
specified in Tab.~\ref{tab::exp} provide an excellent approximation to
the full result. Thus, it can be assumed that the corresponding terms at
three loops approximate the unknown result with high precision.

\section{\label{sec::3loops} Technical details to the three-loop calculation}

The three-loop calculation of the individual Green's functions 
contributing to $M_h$ is organized as
follows: All Feynman diagrams are generated with {\code
  QGRAF}~\cite{Nogueira:1991ex}. In order to 
properly take into account the Majorana character of the gluino, the
output is subsequently manipulated by a {\code PERL}
script~\cite{Harlander:2009mn}  which
applies the rules given in Ref.~\cite{Denner:1992vza}. The various
diagram topologies are identified and transformed to {\code
  FORM}~\cite{Vermaseren:2000nd} with the help of {\code q2e} and {\code
  exp}~\cite{Harlander:1997zb,Seidensticker:1999bb}.  The program {\code
  exp} is also used in order to apply the asymptotic expansion (see,
e.g., Ref.~\cite{Smirnov:2002pj}) in the various mass hierarchies. The
actual evaluation of the integrals is performed with the package {\code
  MATAD}~\cite{Steinhauser:2000ry}, resulting in an expansion in
$d-4$ for each diagram, where $d$ is the space-time dimension.  

The total number of three-loop diagrams amounts to 6706 and 7670 for the
$\phi_1$ and $\phi_2$ self energies, respectively, and 845 and 982 for
the corresponding tadpole contributions.  The computation of the
off-diagonal matrix element $\Sigma_{\phi_{12}}$ involves 6136
diagrams, and the propagator of the pseudoscalar Higgs another 7670.
The application of the asymptotic expansion significantly enlarges these
numbers leading to about 37\,000 (for (h3)) or even 94\,000 (for (h6))
subdiagrams. Note that there are diagrams where, depending on the
hierarchy, up to 15 subdiagrams have to be considered.  A typical
example is shown in Fig.~\ref{fig::diags}(a).

A subtlety arises from diagrams as the one shown in
Fig.~\ref{fig::diags}(b).  If both the external momentum and the
\epscalar{} mass are set to zero from the beginning, an infra-red
divergence occurs and cancels the ultra-violet divergence of the
integral. In effect, the diagram will be of order $(d-4)$ due to the
\epscalar{} algebra. In order to avoid this, we keep the external
momentum $q$ non-zero, though much smaller than all other scales. The
ultra-violet pole multiplied by the algebraic factor of $(d-4)$ then
produces a finite contribution, while the infra-red divergence leads to
$(d-4)\ln(q^2)$ and vanishes as $d\to 4$.

Instead of the requirement $q\not=0$ one could also introduce a nonzero
mass for the \epscalar{}s in order to regulate the infra-red
divergences. In the final result we again observe that the regulator is
multiplied by an additional factor $(d-4)$ leading to a finite 
result for $\Mes\to 0$. We have checked that the latter prescription
leads to identical results as the one with $q\not=0$.

We refrain from presenting all available analytical results for the
expansion in the various regions. They are implemented in the program
\hthreel{} and thus easily accessible if necessary.  However, for the
convenience of the reader we provide in this Section the result for
(h4), see Eq.~(\ref{eq::hierarchies}), which could be useful for other
applications.  We will present the results expressed in terms of the
\drbar{} parameters $\alpha_s$, $m_t$, $\mstop{1}$, $\mstop{2}$,
$\mgluino$ and $\theta_t$.  The corresponding counterterms can be found
in Appendix~A.

\begin{figure}[t]
  \centering
  \begin{tabular}{cc}
    \includegraphics[width=.45\linewidth]{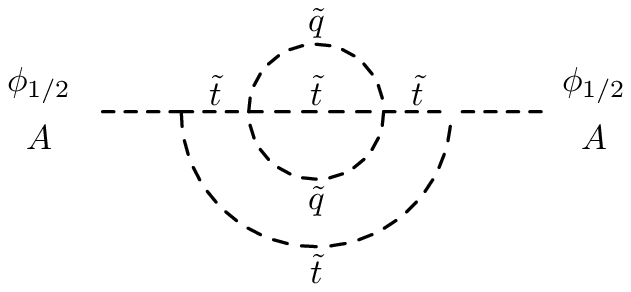}
    &
    \includegraphics[width=.45\linewidth]{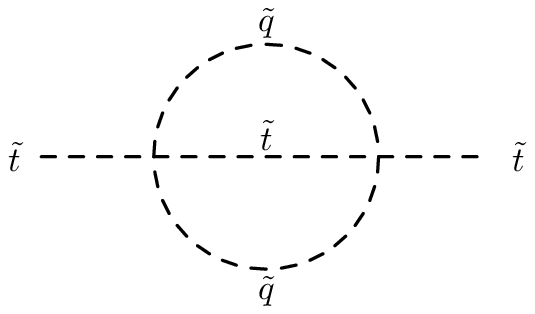}
    \\ (a) & (b)
  \end{tabular}
  \caption{(a) Feynman diagram involving a heavy virtual squark 
    contributing to the Higgs boson self energy. 
    (b) Counterterm diagram related to the diagram in (a).
    The same notation as in Fig.~\ref{fig::diags}
    has been adopted.
  }
  \label{fig::stst_squark}
\end{figure}

Before providing explicit expressions a comment concerning the \drbar{}
renormalization constants for the top squarks is in order.  Due to
diagrams involving heavy squarks $\tilde q$, for example
Fig.~\ref{fig::stst_squark}(a), the squared Higgs boson mass receives
contributions which are proportional to $\msq^2$ and thus can lead to
unnatural large corrections. For this reason we adopt the on-shell
scheme for these contributions to $\delta Z_{{m}_{\tilde{t}_1}}$ and
$\delta Z_{{m}_{\tilde{t}_2}}$ (cf. Eq.~(\ref{eq::mst1}) and
Fig.~\ref{fig::stst_squark}(b) for a sample diagram). This avoids the
potentially large terms $\sim m_{\tilde q}^2$ from the three-loop
diagrams.  We follow this procedure also in the case where the top
squarks and the ``heavy squarks'' are degenerate in mass.  The
renormalization of the mixing angle is free of such enhanced
contributions and we can stick to the pure \drbar{} scheme in that case.
We hasten to add that this discussion only concerns the internal
structure of \hthreel{} and has no direct consequences for the user. The
input parameters of \hthreel{} are the \drbar{} ones as they appear,
e.g., in the output of {\code SOFTSUSY}.

As already noticed in Refs.~\cite{Heinemeyer:1998np,Degrassi:2001yf}, a 
similar behaviour is  observed when the gluino  is much heavier than the  
top squarks.  In this case,  the two- and three-loop 
corrections to the Higgs masses computed in the \drbar{} scheme contain 
terms proportional to  $m_{\tilde{g}}$ and $m_{\tilde{g}}^2$. These 
contributions are canceled in the on-shell scheme by the finite parts of the 
relevant counterterms. 
Thus, in order to avoid unnatural large radiative corrections 
to the Higgs masses, we adopt for scenarios with heavy  gluino masses
a modified renormalization scheme for the top squark masses. We call this
scheme ``modified \drbar{}'' (\drbarmod{}) 
and it is characterized by the non minimal renormalization
 of the   top squark masses. The additional finite shifts of top 
squark masses are chosen such that they  cancel the power-like behaviour
of the gluino contributions. Again, the renormalization of the mixing angle
will not be modified as compared to the genuine \drbar{} scheme.

The relevant finite shifts for the scenarios considered in this paper
are explicitly given in Appendix~B. As can be noticed from
Eq.~(\ref{eq::hierarchies}) the scenarios (h4), (h5) and (h6)
display heavy squark mass contributions  
whereas heavy gluino terms are specific only for the
scenarios (h6) and (h6b). 

In the practical calculation we use the \drbar{} top squark mass parameters as
provided by the spectrum generators and transform them with the help of the
formulae of Appendix~B to the corresponding parameters in the \drbarmod{} scheme
which constitute the input for our analytic expressions.

In the following we present results for the renormalized two-point
functions $\hat\Sigma_{\phi_1}$, $\hat\Sigma_{\phi_2}$ and
$\hat\Sigma_{\phi_1\phi_2}$ for the hierarchy (h4) which for equal top
squark and gluino masses take the form\footnote{We only include terms
  up to order $1/m_{\tilde{q}}^2$.}
\begin{align}
  \hat\Sigma_{\phi_2} 
  &= \frac{
    G_F \Mt^4 \sqrt{2} } {\pi^2 \sin^2\beta}
  \bigg[
  \frac{3}{2} \lmmtMS 
  + \afourpi 
  \bigg(
  4 
  + \Big(
  4 
  + 16 \lmumt
  \Big) \lmmtMS 
  + 4 \lmmtMS^2 
  + \frac{A_t}{\Msusy} \Big(
  4 
  + 8 \lmumt 
  + 4 \lmmtMS
  \Big)
  \bigg) 
  \notag\\&
  + \left(\afourpi\right)^{\!2} \negthickspace
  \begin{aligned}[t]
    \bigg\{{}&
    \frac{2764}{9} 
    - \frac{116}{27} \lmumt 
    - \frac{136}{3} \lmumt^2 
    + \bigg(
 -\frac{644}{9} 
    + \frac{164}{3} \lmumt
    \bigg) \lmmtMS^2 
    \notag\\&
    + 24 \lmmtMS^3 
    + \frac{400}{3} \lmmtMsq 
    - \frac{200}{3} \lmmtMsq^2 
    - \frac{20}{3} \lmmtMsq^3 
    - 120 \z{2} 
    - 80 \lmmtMsq \z{2} 
    + \frac{8}{3} \z{3} 
    \notag\\&
    - \bigg(
    \frac{2216}{27} 
    + \frac{644}{9} \lmumt 
    - \frac{328}{3} \lmumt^2 
    - 40 \lmmtMsq 
    - 20 \lmmtMsq^2 
    - 40 \z{2} 
    + 16 \z{3}
    \bigg) \lmmtMS 
    \notag\\&
    + \frac{\msq^2}{\Msusy^2} \bigg(
    -40 
    + 80 \lmumt
    + 80 \lmmtMsq
    + 80 \z{2}
    \biggr)\notag\\&
    + \frac{\Msusy^2 }{\msq^2}
    \begin{aligned}[t]
      \bigg({}&
      \frac{42356}{225} 
      + 8 \lmmtMS^2 
      - \frac{2128}{45} \lmmtMsq 
      - \frac{176}{3} \lmmtMsq^2 
      \notag\\&
      + \Big(
      \frac{3928}{45} 
      + \frac{152}{3} \lmmtMsq
      \Big) \lmmtMS 
      - \frac{400}{3} \z{2}
      \bigg)
    \end{aligned}
    \notag\\&
    + \frac{A_t \Msusy}{\msq^2}
    \begin{aligned}[t]
      \bigg({}&
      -80 
      + \lmmtMS \Big(
      -\frac{320}{9} 
      - \frac{80}{3} \lmmtMsq
      \Big) 
      + \frac{320}{9} \lmmtMsq 
      + \frac{80}{3} \lmmtMsq^2 
      + \frac{160}{3} \z{2}
      \bigg)
    \end{aligned}
    \notag\\&
    + \frac{A_t}{\Msusy}
    \begin{aligned}[t]
      \bigg({}&
      \frac{832}{27} 
      + \frac{728}{27} \lmumt 
      + \frac{200}{3} \lmumt^2 
      + \frac{608}{9} \lmmtMS^2 
      \notag\\&
      + \lmmtMS \Big(
      \frac{1256}{27} 
      + \frac{800}{9} \lmumt 
      - \frac{160}{3} \lmmtMsq
      \Big) 
      \notag\\&
      - \frac{400}{9} \lmmtMsq 
      + 40 \lmmtMsq^2 
      + 80 \z{2} 
      - \frac{212}{3} \z{3}
      \bigg)
    \end{aligned}
    \notag\\&
    + \frac{A_t^2}{\Msusy^2}
    \begin{aligned}[t]
      \bigg({}&
      -\frac{349}{9} 
      + \frac{32}{9} \lmumt 
      + \frac{32}{9} \lmumt^2 
      + \Big(
      \frac{56}{9} 
      + \frac{64}{9} \lmumt
      \Big)  \lmmtMS 
      \notag\\&
      + \frac{32}{9} \lmmtMS^2 
      + \frac{94}{3} \z{3}
      \bigg)
      \bigg\}
      + \mathcal{O}\left(\frac{\Msusy^4}{\msq^4}\right)
      \bigg]\,,
    \end{aligned}
  \end{aligned}
  \displaybreak[2]\\
  \hat\Sigma_{\phi_1}
  &= \frac{G_F \Mt^4 \sqrt{2}}{\pi^2 \cos^2\beta}
  \begin{aligned}[t]
    \left(\afourpi\right)^{\!2}
    &
    \frac{A_t^2}{\Msusy^2}
    \bigg[
    -\frac{349}{9} 
    + \frac{32}{9} \lmumt 
    + \frac{32}{9} \lmumt^2 
    \notag\\&
    + \bigg(
    \frac{56}{9} 
    + \frac{64}{9} \lmumt
    \bigg) \lmmtMS 
    + \frac{32}{9} \lmmtMS^2 
    + \frac{94}{3} \z{3}
    + \mathcal{O}\left(\frac{\Msusy^4}{\msq^4}\right)
    \bigg]\,,
  \end{aligned}
  \displaybreak[2]\\
  \hat\Sigma_{\phi_{12}}
  &= \frac{G_F \Mt^4 \sqrt{2}}{\pi^2 \cos\beta \sin\beta}
  \bigg[
  \afourpi \frac{A_t}{\Msusy}
  \Big(
  -2 
  - 4 \lmumt 
  - 2 \lmmtMS
  \Big)
  \notag\\&
  + \left(\afourpi\right)^{\!2} \negthickspace
  \begin{aligned}[t]
    \bigg\{{}&
    \frac{A_t^2}{\Msusy^2}
    \begin{aligned}[t]
      \bigg({}&
      \frac{349}{9} 
      - \frac{32}{9} \lmumt 
      - \frac{32}{9} \lmumt^2 
      + \Big(
      -\frac{56}{9} 
      - \frac{64}{9} \lmumt
      \Big) \lmmtMS 
      - \frac{32}{9} \lmmtMS^2 
      - \frac{94}{3} \z{3}
      \bigg)
    \end{aligned}
    \notag\\&
    + \frac{A_t \Msusy}{\msq^2}
    \bigg(
    40 
    - \frac{160}{9} \lmmtMsq 
    - \frac{40}{3} \lmmtMsq^2 
    + \lmmtMS \Big(
    \frac{160}{9} 
    + \frac{40}{3} \lmmtMsq
    \Big) 
    - \frac{80}{3} \z{2}
    \bigg)
    \notag\\&
    + \frac{A_t}{\Msusy} \bigg(
    -\frac{416}{27} 
    - \frac{364}{27} \lmumt 
    - \frac{100}{3} \lmumt^2 
    - \frac{304}{9} \lmmtMS^2 
    + \frac{200}{9} \lmmtMsq 
    - 20 \lmmtMsq^2 
    \notag\\&
    + \lmmtMS \Big(
    -\frac{628}{27} 
    - \frac{400}{9} \lmumt 
    + \frac{80}{3} \lmmtMsq
    \Big) 
    - 40 \z{2} 
    + \frac{106}{3} \z{3}
    \bigg)
    \bigg\}
    + \mathcal{O}\left(\frac{\Msusy^4}{\msq^4}\right)
    \bigg]\,,
  \end{aligned}
\end{align}
with $m_t=m_t(\mu_r)$, $\Msusy=\Msusy(\mu_r)=
\mstop{1}(\mu_r)=\mstop{2}(\mu_r)=\mgluino(\mu_r)$,
$\lmumt=\ln(\mu_r^2/m_t^2)$, $\lmmtMS=\ln(m_t^2/m_{\rm SUSY}^2)$ and
$\lmmtMsq=\ln(m_t^2/m_{\tilde{q}}^2)$ where $\mu_r$ is the
renormalization scale.  The on-shell result corresponding to
$\hat\Sigma_{\phi_2}$ has been presented in Ref.~\cite{Harlander:2008ju}
for $A_t=0$.

We refrain from providing more analytic results since all of them come along
with the program \hthreel{} which is discussed in the next section.

\section{\label{sec::h3l}Description of \hthreel}

In this Section we describe the implementation of our three-loop results
in a user-friendly computer program which allows the evaluation of the
light CP even Higgs boson mass $M_h$ to three-loop accuracy.  The
program is implemented in the form of a {\code Mathematica} package.

To set the input parameters for the calculation, i.e. the SUSY spectrum
and SM parameters, the SUSY Les Houches Accord ({\sc
  slha})~\cite{Skands:2003cj} is used.  For ease of use, we provide functions  
that call a spectrum generator from {\code Mathematica} to produce
an {\sc slha} spectrum file.  To produce the plots in this publication,
we have chosen {\code SOFTSUSY}~\cite{Allanach:2001kg}, but it is possible
to use any spectrum generator that provides the \drbar{} parameters in
addition to the on-shell mass spectrum like {\code SuSpect}~\cite{Djouadi:2002ze} or
{\code SPheno}~\cite{Porod:2003um}. The advantage of {\code SOFTSUSY} is that the
renormalization scale of the \drbar{} parameters can be chosen independently
of the electroweak symmetry breaking.

The corrections $\Delta^{m_t^4} M_h$ to $M_h$, being proportional to the
fourth power of the mass of the top quark $m_t$, are very sensitive to
both the definition and the uncertainty of $m_t$.  Thus, it is
important to use the most precise value of $m_t$ available. For this reason
we take into account the full two-loop SQCD corrections
between the on-shell and \drbar{} top quark mass given in
Ref.~\cite{Martin:2005ch}.\footnote{We thank Steven Martin for providing
  us with the relevant formulae from Ref.~\cite{Martin:2005ch} in electronic
  form, and for allowing us to include his code in our program.}  In
this paper, the relation of the on-shell top mass $M_t$ and \drbar{} top
mass $m_t$ was derived as a function of the \drbar{} masses.  Solving
this equation iteratively, we get $m_t$ as a function of $M_t$.  The
integrals appearing in~\cite{Martin:2005ch} are evaluated using the {\code
  C} library {\code TSIL}~\cite{Martin:2005qm}. This relation is available
for general renormalization scale $\mu_r$ which enables us to obtain
$m_t(\mu_r)$ in the \drbar{} scheme using the on-shell mass $M_t$ as
measured at the Tevatron~\cite{:2009ec} as input.

Another critical parameter for the evaluation of $M_h$ is the strong
coupling $\alpha_s$. We use $\alpha_s(M_Z)$ as input and follow
Ref.~\cite{Harlander:2005wm,Harlander:2007wh,Bauer:2008bj} in order to
evaluate $\alpha_s$ in the \drbar{} scheme with all SUSY particles
contributing to the running.  First, $\alpha_s^{(5),\overline{\rm
    MS}}(M_Z)=0.1184$~\cite{Bethke:2009jm} is run up to the decoupling
scale, which we set to the average value of the SUSY particles, using
the four-loop $\beta$ function~\cite{vanRitbergen:1997va,Czakon:2004bu}.
There, we perform the transition to the \drbar{} scheme and the full
theory.  The two-loop matching coefficients from~\cite{Bauer:2008bj} are
used in this step.  To obtain $\alpha_s^{\rm(full),\overline{\rm
    DR}}(\mu_r)$ for arbitrary values of the renormalization scale
$\mu_r$, we use the three-loop SQCD $\beta$ function given
in~\cite{Jack:1996vg,Harlander:2009mn}.

The remaining input parameters comprise the ones for the SUSY breaking
scenario, which we summarise in the table of Appendix~C
and the SM parameters $M_Z$, $G_F$ and $\alpha$
which also serve as input for the spectrum. The default values set in
\hthreel{} read
\begin{eqnarray}
  M_Z &=& 91.1876~\mbox{GeV}\,,\nonumber\\ M_t &=&
  173.1~\mbox{GeV}\,,\nonumber\\ G_F &=& 1.16637\cdot
  10^{-5}~\mbox{GeV}^{-2}\,,\nonumber\\ 1/\alpha(M_Z) &=&
  127.934\,,\nonumber\\ \alpha_s(M_Z) &=& \alpha_s^{(5),\overline{\rm
      MS}}(M_Z) \,\,=\,\, 0.1184\,.
  \label{eq::para}
\end{eqnarray}
Of course, it is possible to modify these default values.
Note that the top squark masses and mixing angle
are obtained from the soft breaking parameters according to 
Eqs.~(\ref{eq:mst12}) and~(\ref{eq::Atmu}).

In order to include all the known corrections to $M_h$ at the one- and
two-loop level, the spectrum file is passed to {\code
  FeynHiggs}~\cite{Frank:2006yh,Heinemeyer:1998yj,
Degrassi:2002fi,Heinemeyer:1998np}.
{\code FeynHiggs} uses on-shell parameters that are given in the
spectrum file and provides the neutral Higgs mass matrix up to the
two-loop level. Since we prefer to use the \drbar{} scheme, we need to
perform a conversion before adding our three-loop results.  This we do
by subtracting the on-shell expression for $\Delta^{m_t^4} M_h$ (up to
two loops, without any expansions in the masses\footnote{We thank Pietro
  Slavich for sending us the compact formulae from
  Ref.~\cite{Degrassi:2001yf} in electronic form.}) and adding it back
in the \drbar{} scheme.  Thus, we use the \drbar{} scheme to evaluate
$\Delta^{m_t^4} M_h$, which are dominant and sensitive to the top quark
mass, and the on-shell scheme for $\Delta^{\rm rem} M_h$.

The next step is to choose a suitable mass hierarchy for the expansion
of the three-loop corrections. This is done by comparing, at the
two-loop level, the full result from Ref.~\cite{Degrassi:2001yf} with
the expansions in all the mass hierarchies and choosing the one
minimizing the error.  Finally, the three-loop corrections are added,
the neutral Higgs mass matrix is diagonalized, and the mass of the light
Higgs is returned to the user.

\begin{figure}[t]
  \begin{center}
    \includegraphics[width=.75\linewidth]{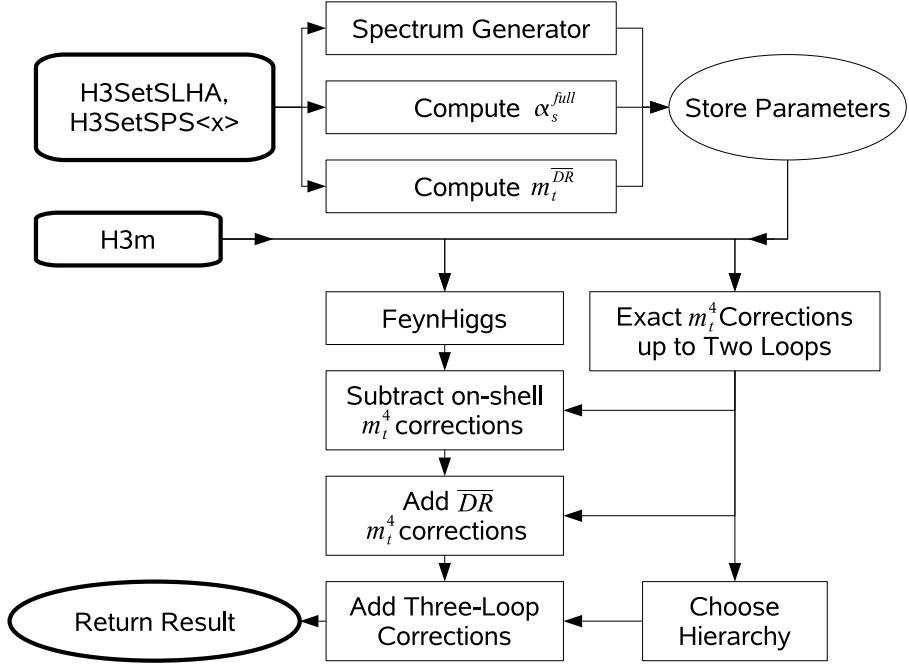}
  \end{center}
  \caption{Flowchart of \hthreel.  First, the user calls {\code H3SetSLHA}
    or one of its descendants to set the parameters.  A subsequent call
    to \hthreel computes $M_h$.
  }
  \label{fig:flow}
\end{figure}

The interface of the program is outlined in Fig.~\ref{fig:flow}.  The
parameters are set up by a call of the function {\code H3SetSLHA}, which
passes its arguments to the spectrum generator and parses its output to
get the relevant input parameters for the calculation.  The top mass and
strong coupling are calculated as described above.  Alternatively, the
function {\code H3GetSLHA} uses an existing spectrum file instead of
running a generator.  For the user interested in the Snowmass Points and
Slopes, we provide convenient wrapper functions {\code H3SetSPS<x>} which
call {\code H3SetSLHA} with parameters according to a specific benchmark
scenario.

The main calculation is organized by the function \hthreel{}, which calls
{\code FeynHiggs}, does the conversion to the \drbar{} scheme described
above, chooses an appropriate mass hierarchy, adds the three-loop
corrections, and returns $M_h$.

In Fig.~\ref{fig:hthreelsession}, a typical {\code Mathematica} session
with \hthreel{} is shown. A more detailed description of \hthreel{} comes
along with the program which can be found at the web page~\cite{h3m}.

\begin{figure}[t]
\begin{verbatim}
Mathematica 7.0 for Linux x86 (64-bit)
Copyright 1988-2008 Wolfram Research, Inc.

In[1]:= Needs["H3`"];

RunDec: a Mathematica package for running and decoupling of the
        strong coupling and quark masses
by K.G. Chetyrkin, J.H. Kuhn and M. Steinhauser (January 2000)

In[2]:= H3SetSPS1a[ 300.];

H3GetSLHA::TSIL: Using TSIL by S.P. Martin.
 -----------------------------------------------------
 FeynHiggs 2.6.5
 built on Dec 20, 2008
 T. Hahn, S. Heinemeyer, W. Hollik, H. Rzehak, G. Weiglein
 http://www.feynhiggs.de
 -----------------------------------------------------
 FHHiggsCorr contains code by:
 P. Slavich et al. (2-loop rMSSM Higgs self-energies)
Loading Results for hierarchy h3
Loading Results for hierarchy h3
Loading Results for hierarchy h6b2qg2
Loading Results for hierarchy h6b2qg2

In[3]:= H3m[]

Loading Results for hierarchy h6b2qg2

Out[3]= {mh -> 114.176}
\end{verbatim}
\caption{A typical {\code Mathematica} session with \hthreel.}
\label{fig:hthreelsession}
\end{figure}

\section{\label{sec::phen}The Higgs boson mass to three-loop accuracy}

In this Section we use, if not stated otherwise, the input parameters
as listed in Eq.~(\ref{eq::para}) and furthermore adopt for the
renormalization scale $\mu_r=\mtop$ as our default value.

\begin{figure}[t]
  \centering
  \begin{tabular}{cc}
    \includegraphics[width=.45\textwidth]{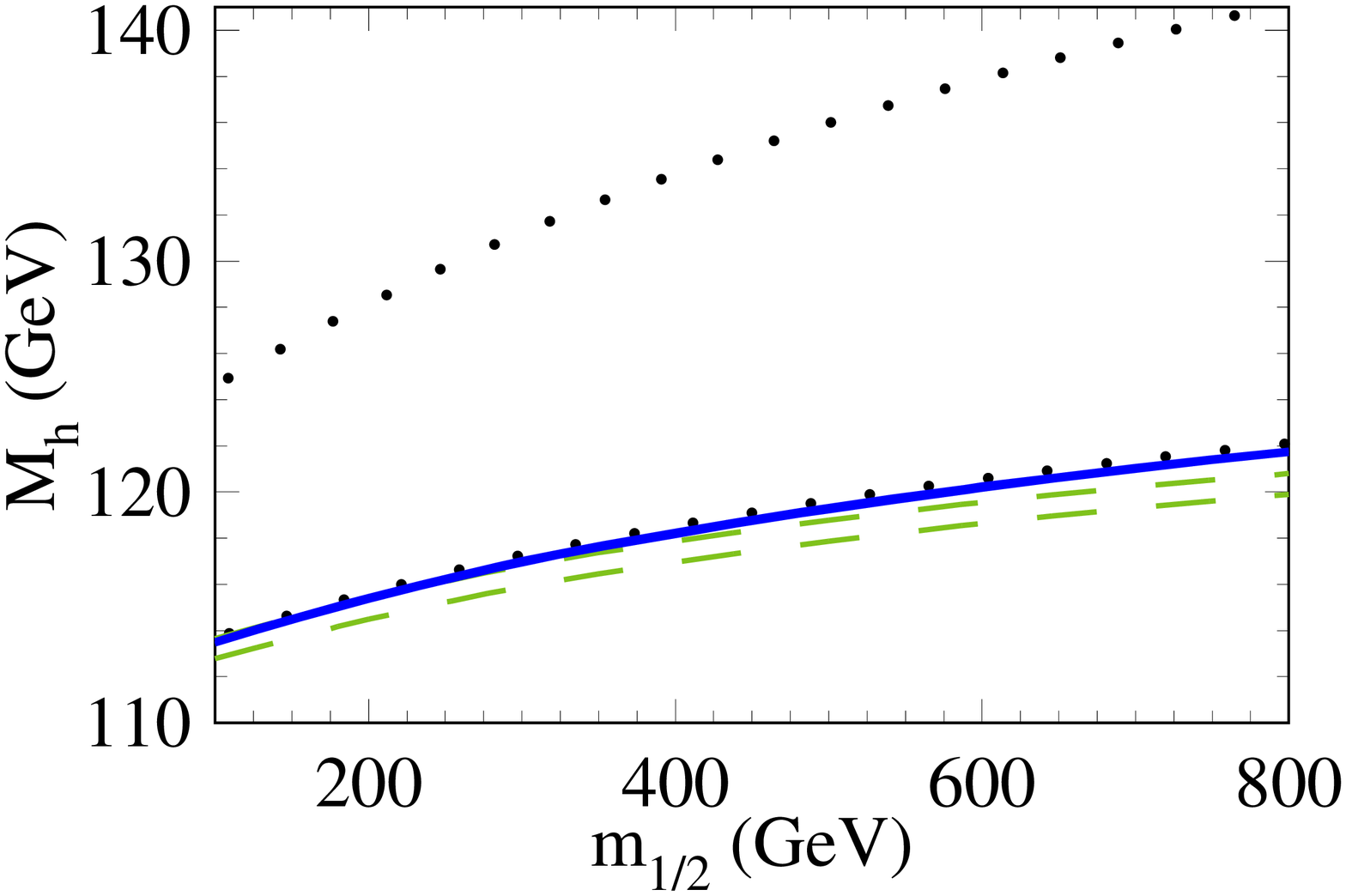}
    &
    \includegraphics[width=.45\textwidth]{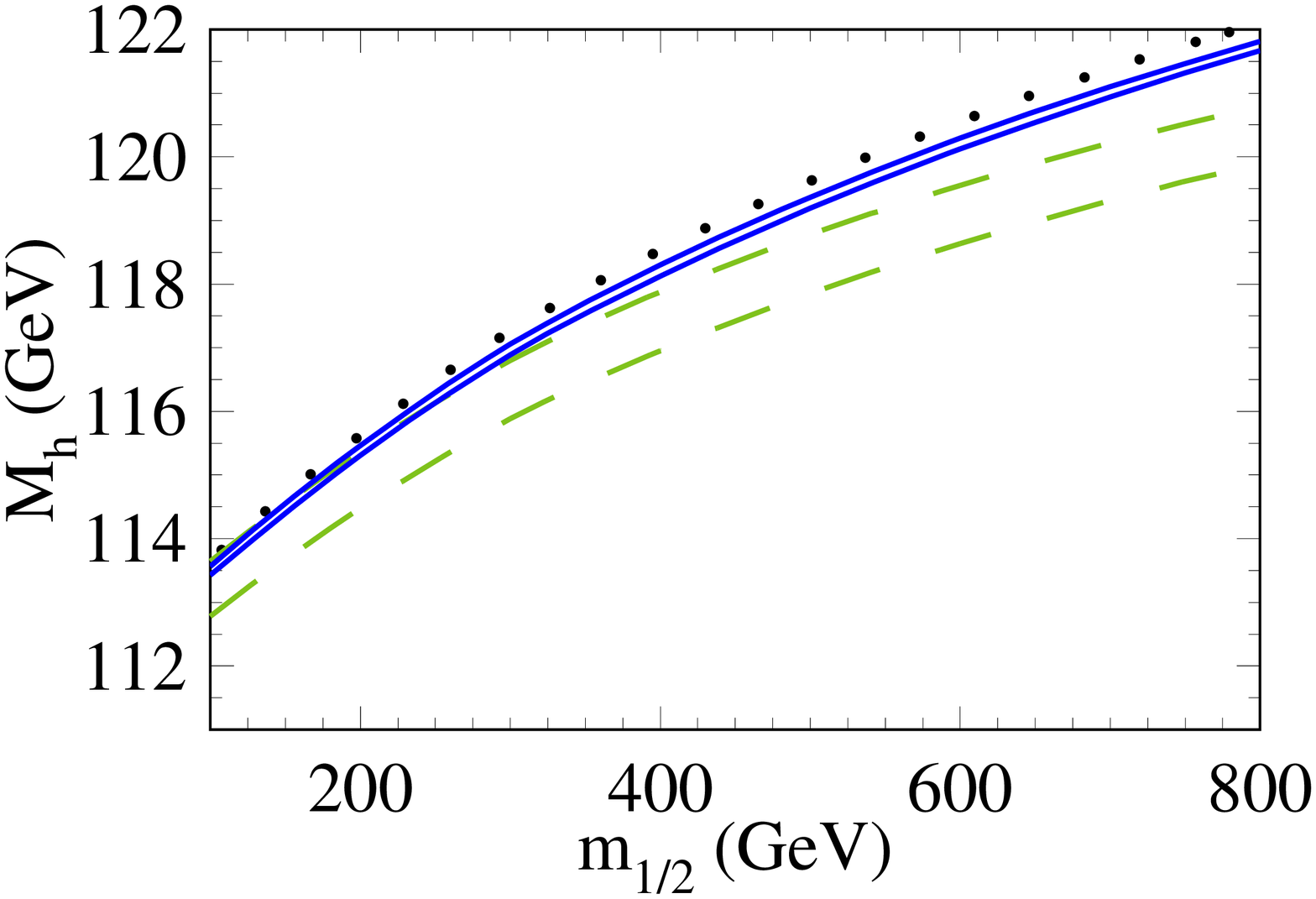}
  \end{tabular}
  \caption{\label{fig::DRvsOS} Renormalization scheme dependence 
    of $M_h$ as a function of $m_{1/2}$ adopting SPS2.
    Dotted, dashed and solid curves correspond to one-, two- and
    three-loop results. The \drbar{} (on-shell) results correspond to the 
    lower (upper) three curves.
    In the right panel the interesting part of the left one is magnified. 
  }
\end{figure}

In Fig.~\ref{fig::DRvsOS} we show the renormalization scheme dependence
of $M_h$ as a function of $m_{1/2}$ for the SPS2 scenario.
This is convenient since we have the same abscissa both for the on-shell and
\drbar{} result.
Note, however, that the three-loop on-shell result is only available for a 
degenerate mass spectrum of the SUSY particles and vanishing parameter
$A_t$~\cite{Harlander:2008ju}. 
Thus, we restrict ourselves to this limit also for the \drbar{} result.
For this reason the following discussion should be considered in a less
quantitative but more qualitative sense and should not be used, e.g., for
estimating a theoretical uncertainty.

In the left panel of Fig.~\ref{fig::DRvsOS}  the upper dotted, dashed and
solid curve correspond to the one-, two- and three-loop prediction of $M_h$ in
the on-shell scheme whereas the corresponding lower three curves 
are obtained in the \drbar{} scheme.
In the on-shell scheme one observes large positive one-loop
corrections which get reduced by 10 to 20~GeV after including the two-loop
terms. The three-loop corrections amount to several hundred MeV. They are
positive or negative --- depending on the value of $m_{1/2}$.

The situation is completely different for \drbar{} mass
parameters: the one-loop corrections are significantly smaller and lead
to values of $M_h$ which are already of the order of the two- and
three-loop on-shell prediction. The two-loop term leads to a small shift
of the order of $-1$~GeV and the three-loop term to a positive shift of
about the same order of magnitude. The final prediction for $M_h$ is
very close to the one obtained after incorporating three-loop on-shell
results.\footnote{The relatively large three-loop corrections (as
  compared to the two-loop ones) do not pose any problem since we use
  simplified formulae as mentioned before. Furthermore, there are
  regions in the parameter space where the two-loop corrections are
  accidentally small in the \drbar{} scheme leading to large
  relative three-loop terms. Nevertheless the overall size of the two-
  and three-loop corrections is small.}

Comparing the \drbar{} and on-shell results in
Fig.~\ref{fig::DRvsOS} one observes a nice reduction of the scheme dependence
when incorporating higher order corrections.\footnote{Up 
  to two-loop order the scheme dependence has already been discussed in 
  Fig.~\ref{fig::mt4}.}
Whereas there is a huge gap between the two one-loop curves (dotted) the
difference in the two-loop prediction of $M_h$ is below 2~GeV which gets
further reduced by about a factor ten after incorporating the three-loop
corrections to roughly 200~MeV.

\begin{figure}[t]
  \centering
  \begin{tabular}{cc}
    \includegraphics[width=.45\textwidth]{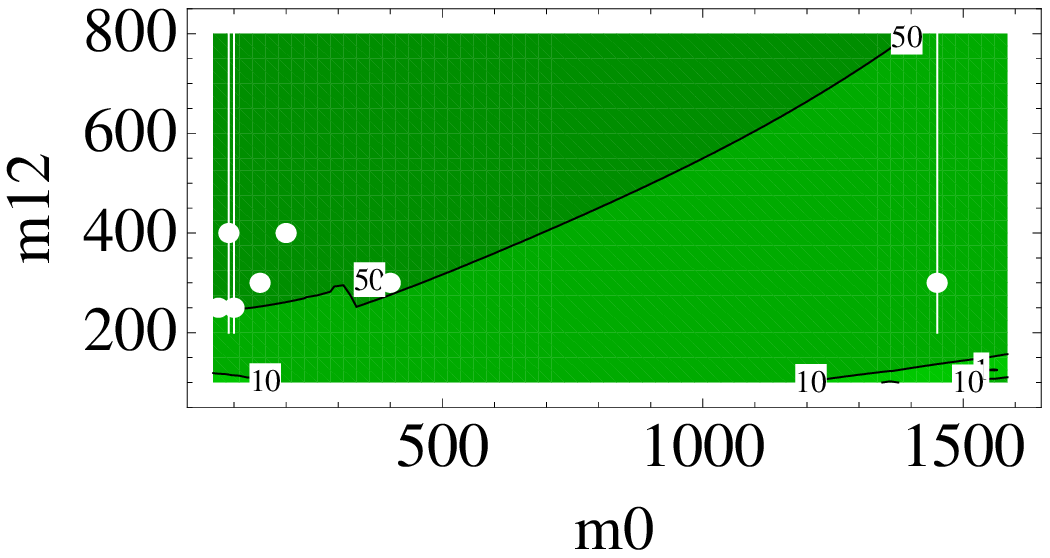}
    &
    \includegraphics[width=.45\textwidth]{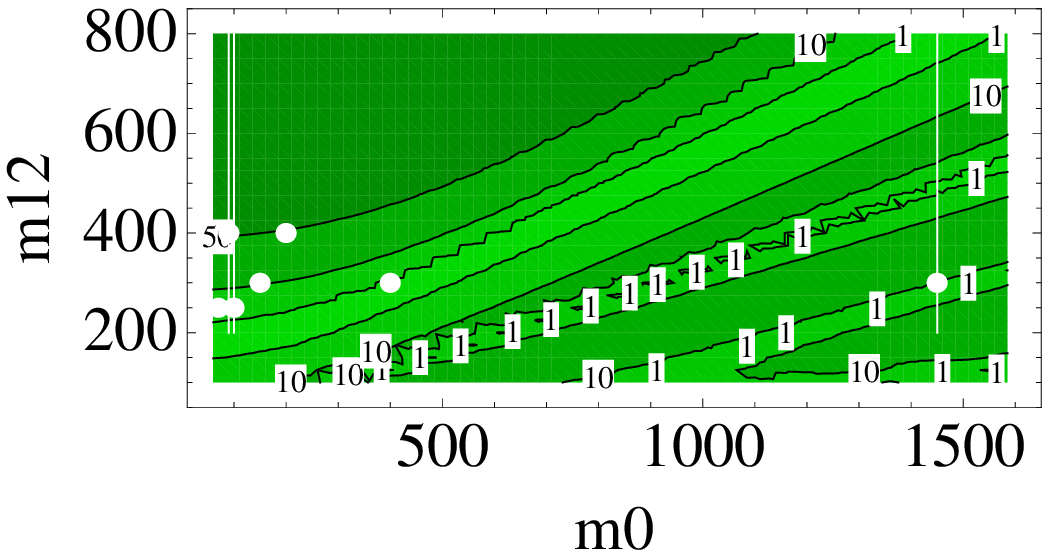}
    \\
    (a) & (b) \\
    \includegraphics[width=.45\textwidth]{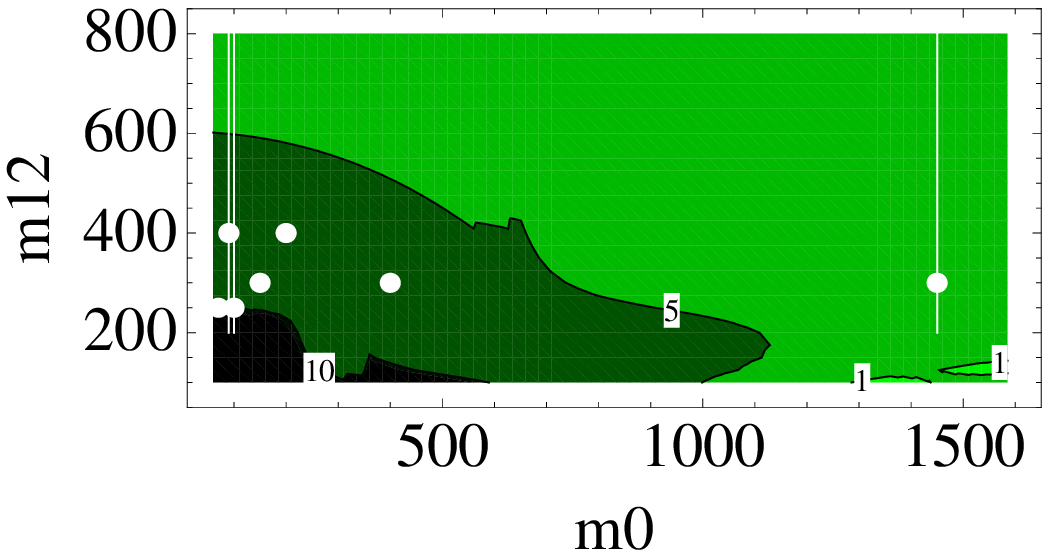}
    &
    \includegraphics[width=.45\textwidth]{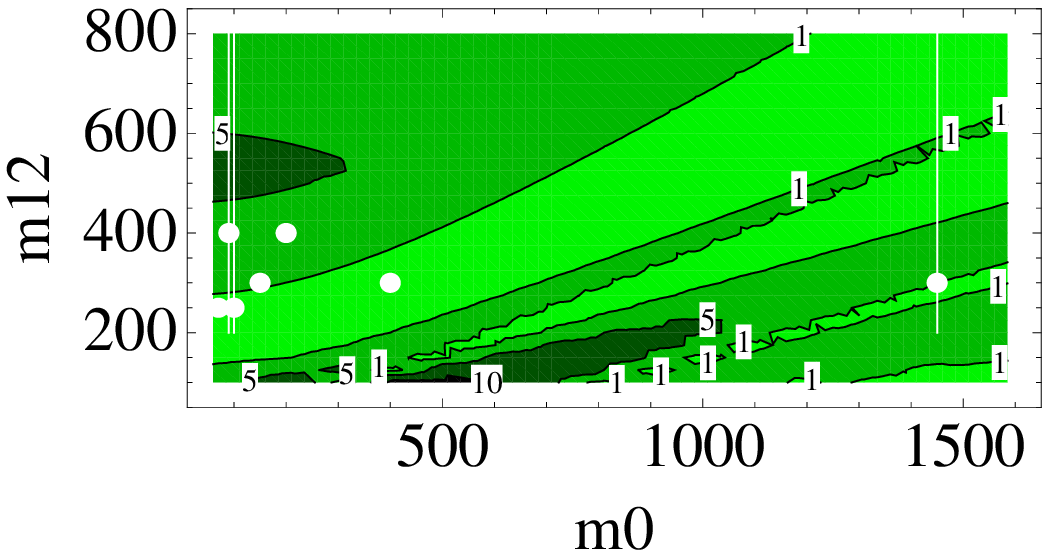}
    \\
    (c) & (d)
  \end{tabular}
  \caption{\label{fig::3lappr_check} Comparison of the three-loop
    predictions for $M_h$ using the maximal available expansion terms
    and reduced input as described in the text. In (a) and (b) the
    absolute deviations are shown for the hierarchy (h3) and the
    combination of (h3), (h5), (h6), (h6b) and (h9), respectively. The
    contour lines indicate the deviations in MeV.  In (c) and (d) the
    results of (a) and (b) are normalized to the genuine three-loop
    contributions where the contour lines indicate the deviations in per
    cent. The benchmark points and slopes are shown as (white) dots and
    lines.  }
\end{figure}

Fig.~\ref{fig::3lappr_check} extends Fig.~\ref{fig::2lappr} to three
loops. In (a) and (b) we again discuss the hierarchy (h3) and the
combination of the hierarchies (h3), (h5), (h6), (h6b) and (h9),
respectively, and show the difference between our best prediction and
the one where the expansion parameters are cut by one unit.\footnote{ We
  only cut in parameters originating from asymptotic expansion: when
  counting powers of mass ratios, we leave $\sin2\theta_t$ untouched and
  do not replace it by Eq.\,(\ref{eq::Atmu}).}  In the whole parameter
plane we observe small absolute corrections reaching at most about
100~MeV.  This leads to the conclusion that as a conservative estimate
of the uncertainty of our approximation procedure one can assign about
100~MeV.

In Fig.~\ref{fig::3lappr_check}(c) and (d) we show relative deviations defined
through 
\begin{eqnarray}
  \delta^{(3)} &=& \frac{M_h^{(3)} - M_h^{(3),\rm cut}}{M_h^{(3)}-M_h^{(2)}}\,,
  \label{eq::delta3}
\end{eqnarray}
where $M_h^{(3)}$ is our best three-loop prediction, $M_h^{(3),\rm cut}$
 is the prediction where at three-loop order the expansion depth of each
 parameter (cf. Tab.~\ref{tab::exp}) is reduced by one unit, and
 $M_h^{(2)}$ corresponds to the (full) two-loop term.  Similar to the
 two-loop case, larger corrections are only observed for small values of
 $m_0$ and $m_{1/2}$ which is a consequence of a small denominator in
 Eq.~(\ref{eq::delta3}).  The three-loop correction terms, however, are
 stable as can be seen from the panels (a) and (b).  Thus, as in the
 two-loop case, we are able to cover the whole $m_0$-$m_{1/2}$ plane and
 are in particular able to produce precise values for $M_h$ for all SPS
 scenarios.

\begin{figure}[t]
  \centering
  \begin{tabular}{cc}
    \includegraphics[width=.45\textwidth]{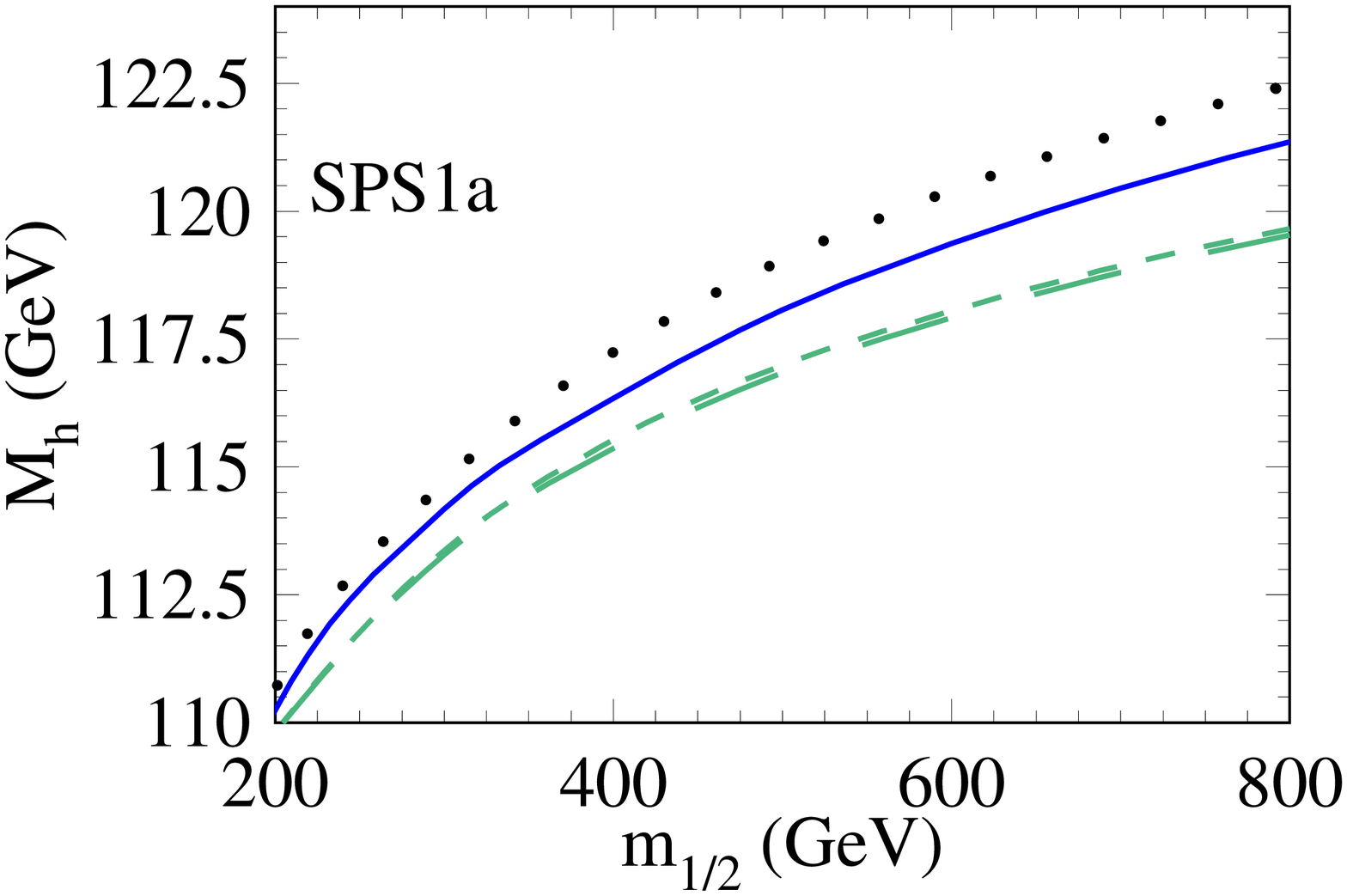}
    &
    \includegraphics[width=.45\textwidth]{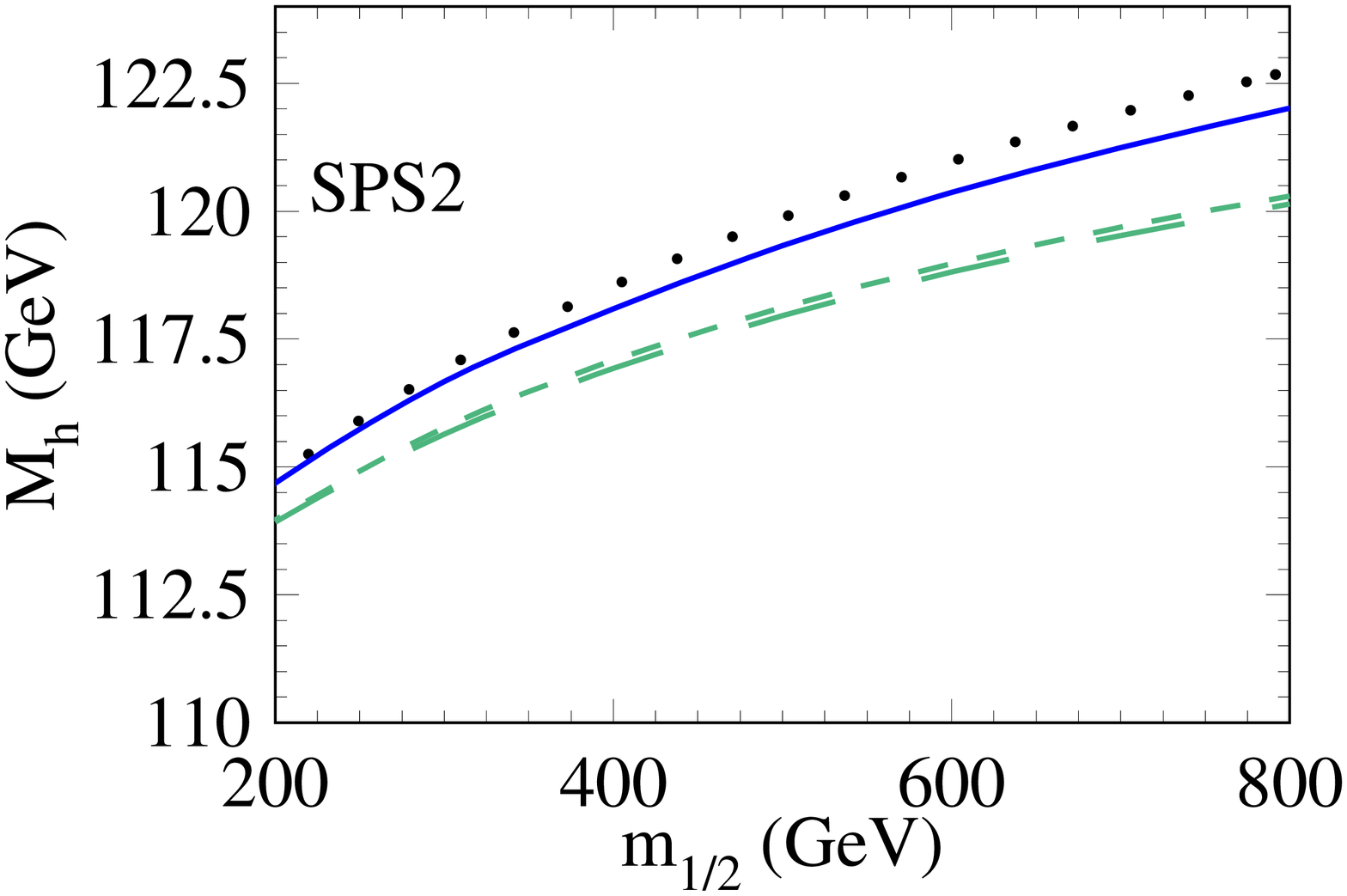}
    \\ (a) & (b) \\
    \includegraphics[width=.45\textwidth]{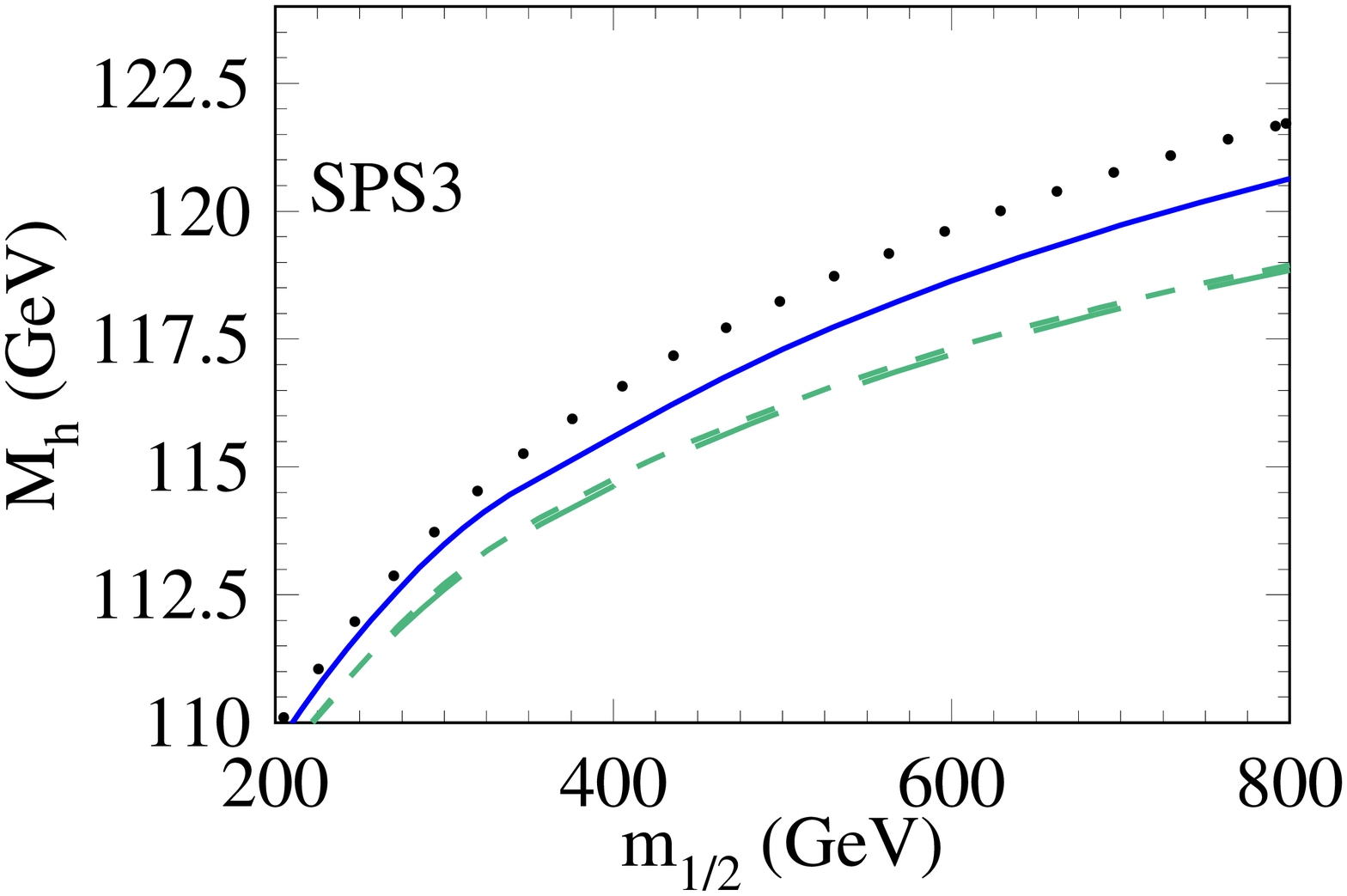}
    &
    \includegraphics[width=.45\textwidth]{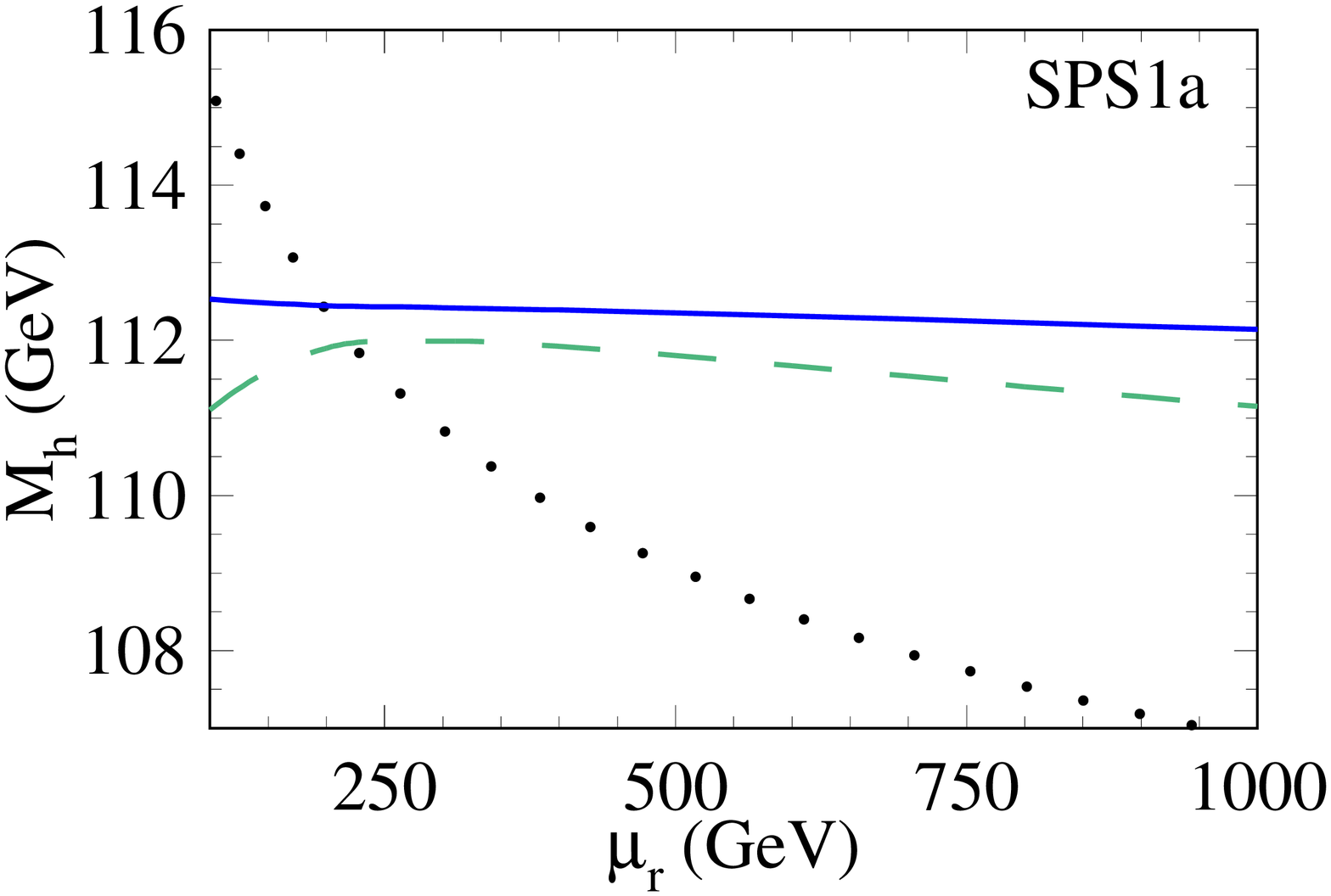}
    \\ (c) & (d)
  \end{tabular}
  \caption{\label{fig::SPS123}$M_h$ for the different slopes of the 
    benchmark scenarios SPS1a (a), SPS2 (b) and SPS3 (c).
    Dotted, dashed and solid lines correspond to the 
    one-, two- and three-loop predictions. 
    The dashed line with 
    longer dashes (at two loops) correspond to the full results, the one with
    the shorter dashes to the approximation implemented in \hthreel{}.
    In (d) the dependence of $M_h$ on the renormalization scale is shown where
    the dotted, dashed and solid line corresponds to the one-, two- and
    three-loop prediction.}
\end{figure}

The slopes for three SPS scenarios are shown in Fig.~\ref{fig::SPS123}(a), (b)
and (c) where
the dotted, dashed and solid lines correspond to the
one-, two- and three-loop predictions, respectively. 
For all three cases one observes negative corrections between 1 and 4~GeV 
at two loops and positive contributions from the three-loop term
which amount up to about 2~GeV.

In Fig.~\ref{fig::SPS123}(d) we show the dependence of the prediction for $M_h$ on
the renormalization scale $\mu_r$. As an example we adopt the SPS1a scenario
with $m_{1/2}=250$~GeV and exploit that {\code SOFTSUSY} allows the
evaluation of all \drbar{} parameters at the scale $\mu_r$.
One observes a strong dependence at one-loop order which gets significantly
reduced at two loops. The three-loop curve is even more flat resulting in a
stable prediction for $M_h$.
Around $\mu_r=250$~GeV the two-loop correction shows a local maximum and is
furthermore very small whereas the three-loop term still amounts to about
500~MeV.
Around $\mu_r=M_t$, which is often used as a default choice, one has negative
two-loop corrections of about $-2$~GeV and a slightly larger three-loop
contribution than for $\mu_r=250$~GeV. 
The corresponding plots for SPS2 and SPS3 look very similar.
Thus, we refrain from presenting them here; they can easily be
generated with the help of \hthreel{}.

\begin{figure}[t]
  \centering
    \includegraphics[width=\textwidth]{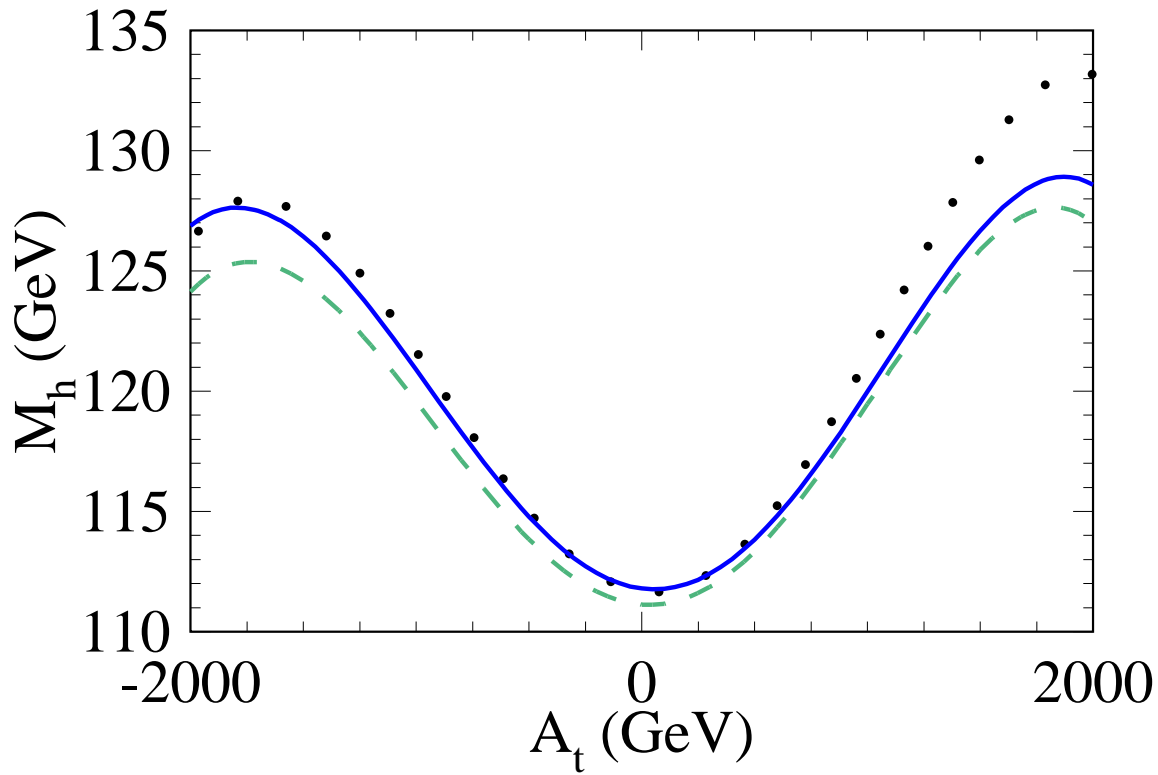}
  \caption{\label{fig::mhAt}Dependence of $M_h$ on $A_t$.
    The dotted, dashed and solid line corresponds to the one-, two- and
    three-loop prediction.
  }
\end{figure}

It is interesting to investigate the dependence of $M_h$ on the soft breaking
parameter $A_t$. In Fig.~\ref{fig::mhAt} we show the result for $M_h$ where
the following values for the parameters have been chosen
\begin{eqnarray}
  \mstop{1} &=& 500~\mbox{GeV}\,,\nonumber\\
  \mstop{2} &=& 1000~\mbox{GeV}\,,\nonumber\\
  \mgluino  &=& 500~\mbox{GeV}\,,\nonumber\\
  \msquark  &=& 2000~\mbox{GeV}\,,\nonumber\\
  \mu_{\rm SUSY}  &=& 800~\mbox{GeV}\,,\nonumber\\
  \tan\beta &=& 10\,,\nonumber\\
  M_A &=& 1500~\mbox{GeV}\,.
  \label{eq::pars1}
\end{eqnarray}
Furthermore, we employ only the $m_t^4$ corrections since it is not possible
to transmit the parameters of Eq.~(\ref{eq::pars1}) directly to {\code
  FeynHiggs} and evaluate the 
corresponding Higgs boson mass.

It is interesting to note that the three-loop corrections are quite
sizeable, amounting up to about 3~GeV. In contrast to the two-loop terms
they are positive and lead to a compensation. For $A_t= -2$\,TeV and
$A_t= 0$ the three-loop prediction is even above the one-loop value for
$M_h$.

\begin{figure}[t]
  \centering
  \begin{tabular}{c}
    \includegraphics[width=\textwidth]{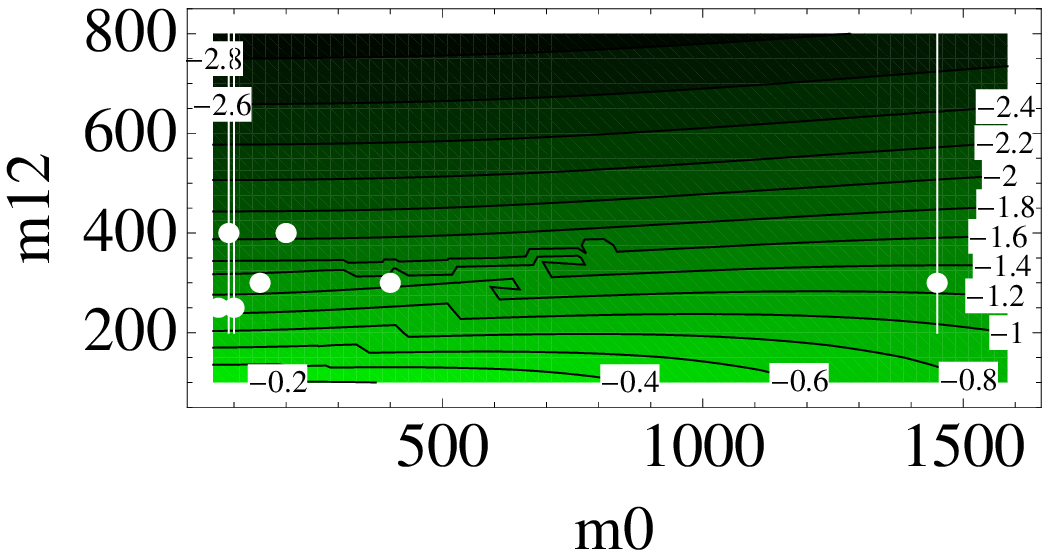}
    \\
    \includegraphics[width=\textwidth]{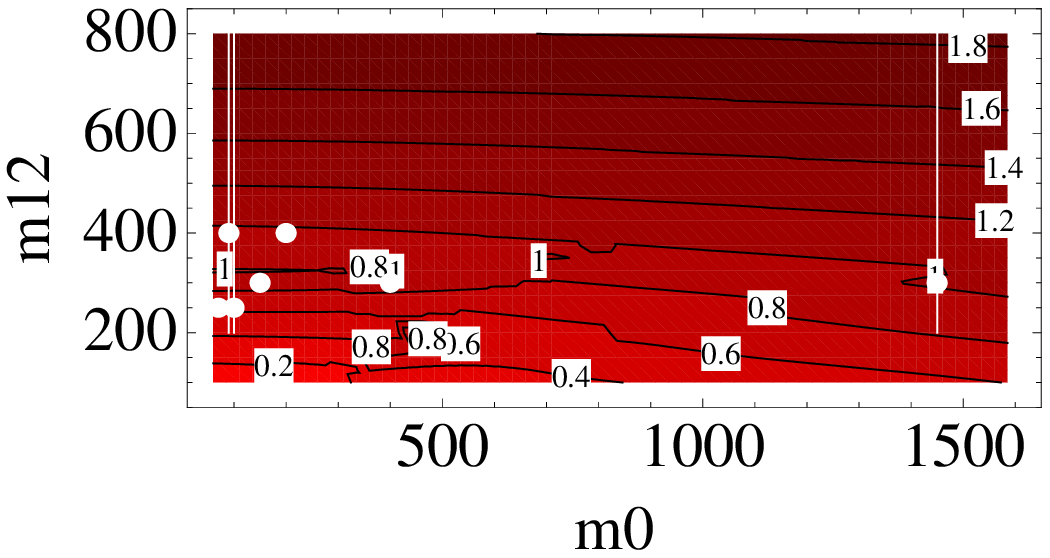}
  \end{tabular}
  \caption{\label{fig::del3l} Genuine two- (upper panel) and three-loop
    (lower panel) corrections to $M_h$ in the $m_0$-$m_{1/2}$ plane.  }
\end{figure}

\begin{figure}[t]
  \centering
  \begin{tabular}{c}
    \includegraphics[width=\textwidth]{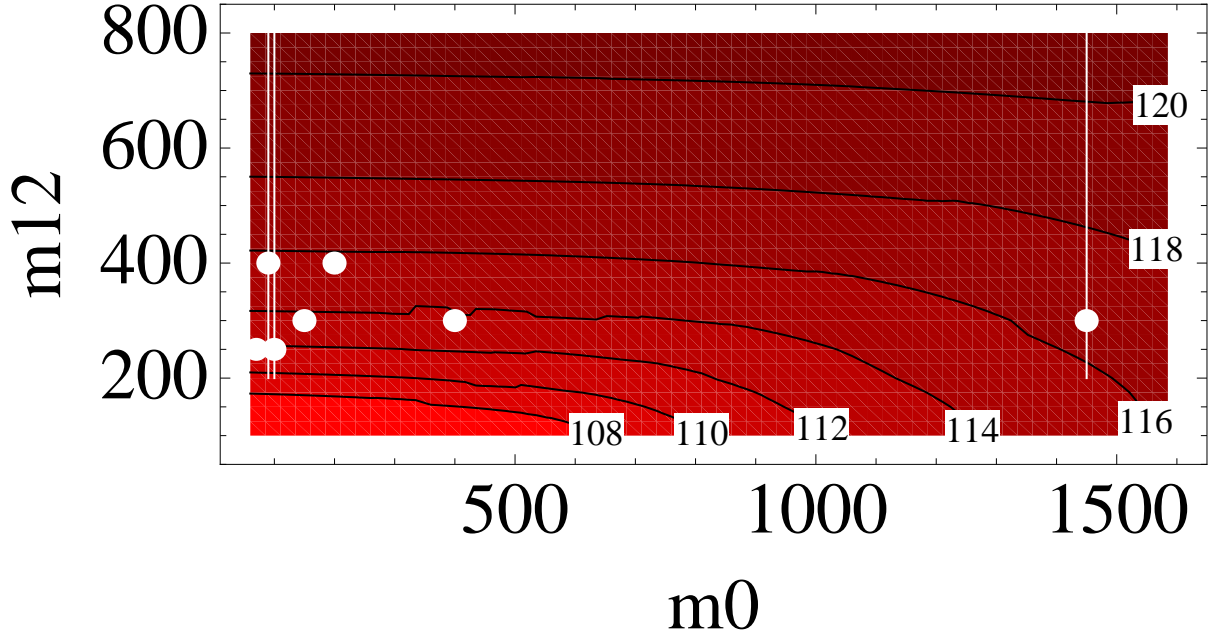}
  \end{tabular}
  \caption{\label{fig::mh3l}
    Prediction of $M_h$ to three-loop accuracy using \hthreel{}.
    The same conventions as in Fig.~\ref{fig::del3l} have been adopted.    
  }
\end{figure}

In order to get an impression on the size of the three-loop corrections
we show in Fig.~\ref{fig::del3l} (lower panel) the difference between
our best three-loop prediction and the full two-loop result as a
function of $m_0$ and $m_{1/2}$.  We observe that the corrections are
always positive and vary for our parameters between a few hundred MeV
and about 2~GeV.  They show only a mild dependence on $m_{0}$, but vary
strongly with $m_{1/2}$. In particular the corrections become larger for
increasing values of $m_{1/2}$.  For comparison, we show in
Fig.~\ref{fig::del3l} (upper panel) the corresponding quantity at
two-loop order, i.e., the difference between the two-loop and the
one-loop result.  In contrast to the three-loop contributions they are
negative and amount to about twice the three-loop terms in a large
region of the parameter space.  However, there are also regions where
the three-loop corrections are larger than the two-loop ones. Note,
however, that in the whole $m_{0}$-$m_{1/2}$ plane the one-loop
corrections are more than ten times bigger.  Furthermore, the occurrence
of three-loop corrections, which are large compared to the two-loop
ones, can also be seen from Fig.~\ref{fig::SPS123}(d): whereas the
two-loop corrections vary between $-4$~GeV and $+4$~GeV in the
considered range of $\mu_r$ the three-loop term is almost constant and
amounts to about 1~GeV relative to the two-loop result.

In Fig.~\ref{fig::mh3l} we finally show the three-loop prediction of
$M_h$ including the three-loop effects discussed in this paper. Again we
restrict ourselves to the parameter space defined above. Values for
other input parameters are easily obtained with the help of \hthreel{}.
One can see that for increasing $m_{1/2}$ also the Higgs boson mass gets
larger and values well above 120~GeV can be reached. This is already
observed at one-loop order and is due to the fact that for larger values
of $m_{1/2}$ the whole supersymmetric spectrum becomes heavier.  For
definiteness we show in Tab.~\ref{tab::SPS3l} the three-loop prediction
of $M_h$ for the SPS points. We show the best prediction ($M_h^{(3)}$)
and, for comparison, the result obtained by cutting parameters as for
Fig.~\ref{fig::3lappr_check} ($M_h^{(3),\rm cut}$). Furthermore, we
indicate the optimal hierarchy as chosen by \hthreel{}.  In all cases
the uncertainty due to our approximation can be estimated to be below
100~MeV.

\begin{table}[t]
  \begin{center}
    \begin{tabular}{c|l|l|c}
      & $M_h^{(3)}$ & $M_h^{(3),\rm cut}$ & optimal \\
      & (GeV)       & (GeV)               & hierarchy \\
      \hline
SPS1a &  112.46 &  112.45 & h6b\\
SPS1a$^\prime$ &  113.92 &  113.92 & h6b\\
SPS1b &  116.49 &  116.44 & h3\\
SPS2 &  116.67 &  116.61 & h5\\
SPS3 &  115.59 &  115.52 & h3\\
SPS4 &  114.82 &  114.81 & h6\\
SPS5 &  112.02 &  111.92 & h3\\
SPS7 &  113.04 &  113.04 & h3\\
SPS8 &  115.03 &  115.02 & h3\\
    \end{tabular}
    \caption{\label{tab::SPS3l}Comparison of the best three-loop prediction to
      the one where some expansion parameters are cut (see text). The last
      column shows the chosen hierarchy.}
  \end{center}
\end{table}

As a last phenomenological application we consider the benchmark points
identified in Ref.~\cite{Carena:2002qg} in order to perform the MSSM
Higgs boson search at hadron colliders. In Tab.~\ref{tab::h_search_scen}
the one-, two- and three-loop predictions for $M_h$ are shown for the four
scenarios ``$m_h^{\rm max}$'', ``no-mixing'', ``gluophobic'' and
``small $\alpha_{\rm eff}$''. Whereas for ``no-mixing'' significant
three-loop effects are observed more moderate, however, still
important contributions are obtained for the remaining three scenarios.
In Tab.~\ref{tab::h_search_scen} $\tan\beta=10$ and
$M_A=1500$~GeV has been chosen, however, a similar conclusion can be
drawn for other values.

\begin{table}[t]
  \begin{center}
    \begin{tabular}{c|c|c|c}
      scenario & 1 loop & 2 loops & 3 loops \\
      \hline
      $m_h^{\rm max}$          & 141.11 & 133.96 & 134.38 \\
      no-mixing               & 121.33 & 119.72 & 121.07 \\
      gluophobic              & 118.52 & 117.47 & 117.29 \\
      small $\alpha_{\rm eff}$ & 120.20 & 118.33 & 119.10 \\
    \end{tabular}
    \caption{\label{tab::h_search_scen}
      Results for $M_h$ (in GeV) for the benchmark scenarios defined
      in Ref.~\cite{Carena:2002qg} where 
      $\tan\beta=10$ and $M_A=1500$~GeV has been chosen.
      }
  \end{center}
\end{table}

Let us at the end of this Section estimate the theoretical uncertainty
on the prediction of $M_h$ after including the three-loop
corrections. We divide the uncertainty into two parts: $(i)$ the
theoretical error due to missing higher order corrections and other as
of yet uncalculated corrections, and $(ii)$ the parametric uncertainty,
mainly due to the top quark mass, $\alpha_s$ and the supersymmetric
masses.  At two-loop level,
a thorough investigation of the theoretical
uncertainties (i) has been performed in Ref.~\cite{Allanach:2004rh} (for
earlier work, see Ref.~\cite{Degrassi:2002fi}). 
It was found that missing two-loop corrections (most
importantly electro-weak and finite-momentum effects) can be assumed to
be well below 1~GeV, while the as of then unknown three-loop effects
could amount to 2-3~GeV, as indicated by the variation of the two-loop
results with the renormalization scale.

With the $\alpha_t\alpha_s^2$ corrections of our calculation, we should
therefore be able to reduce the theory uncertainty. Instead of
renormalization scale variations, however, we want to adopt a more
conservative attitude and assume a geometric progression of the
perturbative series. As can be confirmed at two-loop level, a conservative
estimate of the uncertainty is 50\% of the difference to the next lower
order. Thus we assign 50\% of the three-loop contribution to
$M_h$ as a theoretical error. For the {\sc msugra} scenarios this leads
to an uncertainty of about $100-200$~MeV for $m_{1/2}=100$~GeV and to
about 1~GeV for $m_{1/2}=1$~TeV. These
estimates cover also corrections from renormalization group
improvements which at two-loop order lead to shifts in $M_h$ of the
order of a few GeV~\cite{Carena:2000dp}.

Earlier in this section, we have identified two more contributions to
the uncertainties of type (i): The corrections beyond the quartic top
quark mass contribution, $\Delta^{\rm rem}M_h^{(3)}$, has been estimated
to about $100$~MeV and the uncertainty due to our approximation
procedure also amounts to at most $100$~MeV. Both contributions are
smaller than the one due to missing higher order corrections discussed
above.

The parametric uncertainties can be easily estimated with the help of
\hthreel{}. For definiteness we adopt in the following SPS2 and
vary $M_t$ and $\alpha_s(M_Z)$ as follows
\begin{eqnarray}
  M_t &=& 173.1 \pm 1.3~\mbox{GeV}\,,
  \\
  \alpha_s(M_Z) &=& 0.1184 \pm 0.0020\,.
\end{eqnarray}
In the case of the top quark mass we observe for $m_{1/2}=100$~GeV 
a variation of $M_h$ by about 350~MeV which increases to 
$\delta M_h=1$~GeV for $m_{1/2}=1$~TeV. The corresponding numbers for 
the uncertainty in $\alpha_s(M_Z)$ read 80~MeV and 600~MeV.
A further uncertainty is connected to the (unknown)
supersymmetric parameters which may also be of the order of a few
hundred MeV. Assuming, e.g., an uncertainty of 10\% for $\mstop{1}$
in the range between 200 and 800~GeV and adopting the remaining
parameters from Eq.~(\ref{eq::pars1}) leads to an uncertainty of at most
500~MeV for $M_h$.
Note that the parametric uncertainty is of the same order of magnitude as the
theoretical error due to missing higher order corrections as estimated in the
previous paragraph.

\section{\label{sec::concl}Conclusions}

In this paper we have discussed the lightest Higgs boson mass of the
MSSM to three-loop accuracy.  At this order an exact calculation is out
of range and thus we heavily exploit the methods of asymptotic expansion
in order to provide precise approximations. This procedure has been
successfully tested at two-loop order against the full result. The result is
expressed in terms of \drbar{} parameters for the quark and squark
masses for which we have found very good convergence of the perturbative
expansion.  We provide a user-friendly {\code Mathematica} program
\hthreel{} which allows the computation of $M_h$ in a simple way. In
particular, it is possible to apply various SUSY breaking scanarios,
invoke a spectrum generator, and use the output in order to compute
$M_h$.

As already mentioned, with the help of the asymptotic expansion we have
implemented analytical results valid for various hierarchies in the
supersymmetric masses. \hthreel{} is set up in such a way that it is 
straightforward to include further hierarchies in case they are needed for
future investigations.

We have performed several studies with the new three-loop
corrections. In particular, we have considered their renormalization
scheme and scale dependence, and their numerical effect in the various
SPS scenarios. Furthermore, assuming {\sc msugra}, we have considered
the $m_0$-$m_{1/2}$ parameter space and computed two and three-loop
corrections. On the basis of these investigations we estimate the
remaining theory uncertainty of $M_h$ to about 200~MeV for
$m_{1/2}=100$~GeV and to about 1~GeV for $m_{1/2}=1$~TeV.

\section*{Acknowledgements}

We thank Steven Martin for providing us with {\code C++} routines for
the relation between the \drbar{} and on-shell top quark mass, and
Pietro Slavich for providing us with the analytic result for the
two-loop SQCD corrections. We are grateful to Sven Heinemeyer for
carefully reading the manuscript and many valuable comments, and to
Thomas Hermann for useful comments on Appendix~A. This work was
supported by the DFG through SFB/TR~9 ``Computational Particle Physics''
and contract HA~2990/3-1, and by the Helmholtz Alliance ``Physics at the
Terascale''.

\section*{Appendix A: $\overline{\rm DR}$ counterterms}

In this appendix we provide details about the computation of the counterterms
for the gluino, top quark and squark masses and the mixing angle in the top
squark system.

To define our framework for the computation of the top squark masses and
mixing angle counterterms, we 
start from the bare Lagrangian containing the  kinetic energy and
mass terms
\begin{eqnarray}
{\cal L}_{\tilde{t}}^{(0)} = \frac{1}{2}\partial_{\mu}(\stl^\ast,
\str^\ast)^{(0)}\partial^{\mu}
\left(\begin{array}{c}
\stl\\
\str
\end{array}\right)^{\!\!(0)}-\frac{1}{2}(\stl^\ast,
\str^\ast)^{(0)} ( {\cal M}_{{\tilde t}}^{2})^{(0)}
\left(\begin{array}{c}
\stl\\
\str
\end{array}\right)^{\!\!(0)}\,,
\label{eq::lbare}
\end{eqnarray}
where the superscript $(0)$ labels the bare quantities, $\stl$ and $\str$
denote the interaction eigenstates and the  top squark mass matrix 
 was defined in Eq.~(\ref{eq::mstop}).

The  top squark mass eigenstates are  related to the interaction
eigenstates by
\begin{eqnarray}
\left(\begin{array}{c}
\stu\\
\std
\end{array}\right)^{\!\!(0)} = {\cal R}_{{\tilde t}}^{(0)\, \dag}
\left(\begin{array}{c} 
\stl\\
\str
\end{array}\right)^{\!\!(0)}\,,
\end{eqnarray} 
where the matrix ${\cal R}_{{\tilde t}}$ is defined in Eq.~(\ref{eq::mixa}).

The wave function renormalization can be written in matrix form
\begin{eqnarray}
\left(\begin{array}{c}
\stu\\
\std
\end{array}\right)^{\!\!(0)} ={\cal Z}_{\tilde t}^{1/2} \left(\begin{array}{c}
\stu\\
\std
\end{array}\right)\,,\quad\mbox{with} \quad {\cal Z}_{\tilde t}^{1/2}=
\left(\begin{array}{cc} 
Z_{11}^{1/2}& Z_{12}^{1/2}\\
Z_{21}^{1/2}& Z_{22}^{1/2}
\end{array}\right)\,,
\end{eqnarray}
where we have
${\cal Z}_{\tilde t}^{1/2} = \mathbf{I} +{\cal O}(\alpha_s)$. 
Thus, 
$ Z_{11}^{1/2} = 1 + {\cal  O}(\alpha_s)$,
$Z_{22}^{1/2} = 1 + {\cal  O}(\alpha_s)$, 
$Z_{12}^{1/2} = {\cal O}(\alpha_s)$
and $Z_{21}^{1/2} =  {\cal  O}(\alpha_s)$. \\
Similarly, the renormalized mass matrix can be parametrized as follows
\begin{eqnarray}
\left(
\begin{array}{cc}
m_{\tilde{t}_1}^2& 0\\
0& m_{\tilde{t}_2}^2
\end{array}
\right)^{\!\!(0)} =\left( \begin{array}{cc}
m_{11}^2 Z_{m_{11}}& m_{12}^2 Z_{m_{12}}\\
m_{21}^2 Z_{m_{21}}& m_{22}^2 Z_{m_{22}}
\end{array}
\right)\equiv {\cal M}\,,
\label{eq::masren}
\end{eqnarray}
where the $m_{ij}$ stand for renormalized mass parameters.

In order to extract the renormalization constants it is convenient to consider
the renormalized inverse top squark propagator which reads
\begin{eqnarray}
i S^{-1}(p^2) = p^2 ({\cal Z}_{\tilde t}^{1/2})^\dag{\cal Z}_{\tilde
  t}^{1/2}-
({\cal Z}_{\tilde t}^{1/2})^\dag[{\cal M} -\Sigma(p^2) ] {\cal
  Z}_{\tilde t}^{1/2} \,,
\end{eqnarray}
where $\Sigma(p^2)$ denotes the matrix of self energies
 constructed by the $\tilde{t}_1$ and $\tilde{t}_2$ fields.

To determine the renormalization constants introduced above one has to
specify the renormalization conditions. In the \drbar{} scheme the finite parts
of the renormalization constants are set to zero and the coefficients of the
$\epsilon$ poles are obtained from the requirements
\begin{eqnarray}
  i S_{ij}^{-1}(p^2)\bigg|_{\rm pp}  &=& 0\,, \nonumber\\
  m_{12}^2 Z_{m_{12}}\,\,=\,\,
  m_{21}^2 Z_{m_{21}} &=& 0\,,
\label{eq::rencond}
\end{eqnarray}
where ``pp'' stands for the ``pole part''.
The  conditions in the second line of Eq.~(\ref{eq::rencond}) ensure that the
 renormalized fields
$\stu$ and $\std$ are the mass eigenstates of the renormalized mass
matrix. Accordingly, we can identify $ Z_{m_{11}}$ and $Z_{m_{22}}$ with the
\drbar{} renormalization constants $ Z_{m_{\tilde{t}_1}}$ and $Z_{m_{\tilde{t}_2}}$
 defined through
\begin{eqnarray}
(m_{\tilde{t}_i}^2)^{(0)}=m_{\tilde{t}_i}^2 Z_{m_{\tilde{t}_i}}\,,\quad i=1,2\,.
\end{eqnarray}
In the \drbar{} scheme the wave function renormalization
constants are independent of the masses, so we can evaluate them
solving Eqs.~(\ref{eq::rencond}) for
$m_{\tilde{t}_1}=m_{\tilde{t}_2}=0$.
In addition, we can   exploit the symmetry of the matrix ${\cal Z}_{\tilde
  t}^{1/2} $  and  reparametrize it  in terms of the common
wave-function renormalization 
constant of the fields $\stu$ and  $ \std$ and the counterterm of the mixing
angle, as follows
\begin{eqnarray}
{\cal Z}_{\tilde  t}^{1/2} \equiv \tilde{Z}_2^{1/2}
\left(\begin{array}{cc}
\cos \delta\theta_{t} & \sin \delta\theta_{t} \\
-\sin \delta\theta_{t} & \cos \delta\theta_{t}
\end{array}
\right)\,.
\label{eq::param}
\end{eqnarray}  
We solve Eqs.~(\ref{eq::rencond}) iteratively by inserting the
perturbative expansion of all quantities,
\begin{eqnarray}
  Z&=&1+\as\delta Z^{(1)}
  +\left(\as\right)^2\delta Z^{(2)}+{\cal O}(\alpha_s^3)\,,\nonumber\\
  \delta\theta_t&=&\as
  \delta\theta_t^{(1)}+\left(\as\right)^2\delta\theta_t^{(2)}+{\cal 
    O}(\alpha_s^3)\,,\nonumber\\ 
  \Sigma_{ij}&=& \as \Sigma_{ij}^{(1)}+\left(\as\right)^2\Sigma_{ij}^{(2)}+{\cal
    O}(\alpha_s^3)\,,\quad i,j=1,2\,,
\end{eqnarray} 
and take into account the reparametrization of Eq.~(\ref{eq::param}).
At one-loop order we get
\begin{eqnarray}
  \left[ \Sigma_{ii}^{(1)}
    -m_{\tilde{t}_i}^2(\delta Z_2^{(1)}
    +\delta Z_{m_{\tilde{t}_i}}^{(1)})
    +p^2 \delta Z_2^{(1)}\right]\bigg|_{\rm pp}
  &=&0\,,\quad i=1,2\,,
  \nonumber\\
  \left[\Sigma_{12}^{(1)}
    -\delta\theta_{t}^{(1)}(m_{\tilde{t}_1}^2-m_{\tilde{t}_2}^2)
  \right]\bigg|_{\rm pp}
  &=&0
  \,.
\end{eqnarray}
In a first step we solve the first equations with
$m_{\tilde{t}_1}=m_{\tilde{t}_2}=0$ to obtain
\begin{eqnarray}
  \delta Z_2^{(1)} &=&
  -\frac{1}{p^2}\Sigma_{ii}^{(1)}(p)\bigg|_{\rm pp}\,,
  \quad i=1 \,\, \mbox{or}\,\, i=2\,.
\end{eqnarray}
Afterwards, we determine $\delta Z_{m_{\tilde{t}_i}}^{(1)}$ such that
the equations are fulfilled also for finite $m_{\tilde{t}_i}$. The
renormalization conditions for $S_{12}^{-1}$ and $S_{21}^{-1}$ in
Eq.~(\ref{eq::rencond}) provide an expression for
$\delta\theta_t^{(1)}$:
\begin{eqnarray}
  \delta\theta_t^{(1)} &=&
  \frac{\Sigma_{12}^{(1)}}{m_{\tilde{t}_1}^2-m_{\tilde{t}_2}^2}\bigg|_{\rm pp} \,.
\end{eqnarray}
Similarly, at two-loop order we have 
\begin{eqnarray}
  \Bigg[\Sigma_{ii}^{(2)}+\Sigma_{ii}^{(1)} \delta
  Z_2^{(1)}-m_{\tilde{t}_i}^2(\delta Z_2^{(2)}
  +\delta Z_2^{(1)}\delta Z_{m_{\tilde{t}_i}}^{(1)} +\delta
  Z_{m_{\tilde{t}_i}}^{(2)})&&\nonumber\\
  +(-1)^{i+1}\delta\theta_t^{(1)}[-2 \Sigma_{12}^{(1)}+
  \delta\theta_t^{(1)}(m_{\tilde{t}_1}^2-m_{\tilde{t}_2}^2)] +p^2 \delta
  Z_2^{(2)}\Bigg]\bigg|_{\rm pp} &=& 0 \quad (i=1,2)\,,\nonumber\\
  \Bigg[\Sigma_{12}^{(2)}+\Sigma_{12}^{(1)} \delta
  Z_2^{(1)}+\delta\theta_t^{(1)}(\Sigma_{11}^{(1)}-\Sigma_{22}^{(1)}
-m_{\tilde{t}_1}^2\delta
  Z_{m_{\tilde{t}_i}}^{(1)}+m_{\tilde{t}_2}^2\delta
  Z_{m_{\tilde{t}_2}}^{(1)})&&\nonumber\\
  -(m_{\tilde{t}_1}^2-m_{\tilde{t}_2}^2)
  (\delta\theta_t^{(2)}+\delta\theta_t^{(1)}\delta Z_2^{(1)})
  \Bigg]\bigg|_{\rm pp}&=&0\,.
\end{eqnarray}
In the following we list the results for the squark mass counterterms
and the mixing angle, however, we refrain from providing explicit results
 for $ \delta Z_2^{(1)}$ and $ \delta Z_2^{(2)}$, as we do not need them
 for the actual computation.  

Our results for the  renormalization constant of the top  squark mass
$\Mstu$  read
{\allowdisplaybreaks
\begin{align}
  \Mstu^2\delta Z_{{m}_{\tilde{t}_1}}^{(1)} &= \cf
  \left(-\Mgl^2 - \Mt^2 + \Mgl\Mt\Smt +
    \frac{\Mstd^2-\Mstu^2}{4}\Smt^2\right)\frac{1}{\ep}\,,\nonumber\\
  \Mstu^2\delta Z_{{m}_{\tilde{t}_1}}^{(2)} &= \bigg\{
  \cf^2\bigg[\frac{\Cmt^2\Mgl^2\Mt^2}{\dms} + \frac{(1 + \Cmt^2)\Smt^2(\dms) +
    8\Mt^2}{16} -  \frac{(1 + \Cmt^2)\Mgl\Mt\Smt}{2} \bigg] 
\nonumber\\
&+
\ca\cf \bigg[\frac{9\Mgl^2}{8} + 3\frac{\Smt^2(\dms) + 4\Mt^2}{32}
  - \frac{3\Mgl\Mt\Smt}{4}\bigg]
\nonumber\\
&+
\cf\Nf\TF\bigg[\frac{-3\Mgl^2}{4} - \frac{\Smt^2(\dms) + 4\Mt^2}{16} + \frac{\Mgl\Mt\Smt}{2} \bigg]
\bigg\}\frac{1}{\ep^2}
\nonumber\\
&+ \bigg\{
\cf^2\bigg[\frac{3\Mgl^2}{4} + \frac{\Smt^2(\dms) + 4\Mt^2}{16} - \frac{\Mgl\Mt\Smt}{2} \bigg] 
\nonumber\\
&+
\ca\cf \bigg[\frac{-11\Mgl^2}{8} - 3\frac{\Smt^2(\dms) +
    4\Mt^2}{32} + \frac{3\Mgl\Mt\Smt}{4}\bigg] 
\nonumber\\
&+\cf\Nq\TF\bigg[\frac{3\Mgl^2}{4} + \frac{\Smt^2(\dms) + 8\msquark^2
    + 4\Mt^2}{16} - \frac{\Mgl\Mt\Smt}{2}\bigg] 
\nonumber\\
&+\cf\Nt\TF\bigg[
\frac{3\Mgl^2}{4} + \frac{\Smt^2(\dms) + 4\Mstu^2 + 4\Mstd^2 - 4\Mt^2}{16} - 
 \frac{\Mgl\Mt\Smt}{2}
\bigg]
\bigg\}\frac{1}{\ep}
\nonumber\\
&+\Mes^2 \left(-\ca\cf\frac{3}{8} + \cf\Nf\TF\frac{1 }{4} 
\right)\frac{1}{\ep}\,, 
\label{eq::mst1}
\end{align}
}%
where we have introduced the abbreviations $\Nf=\Nt+\Nq$, 
$\Smt=\sin 2 \theta_t$ and $ \Cmt=\cos 2 \theta_t$.
$N_q$ denotes the number of light quark flavours and takes in our case the value
$N_q=5$. $N_t=1$ has been introduced for convenience. Furthermore
   $\Mes$ denotes the
on-shell renormalized \epscalar{} mass. 
The corresponding results for $\Mstd$ can be derived
from Eq.~(\ref{eq::mst1})
by interchanging $\Mstu$ and $\Mstd$ and changing the sign of
$\theta_t$.

We have employed two different approaches for the renormalization of the 
\epscalar{} 
mass. In one approach we renormalize it on-shell and choose  $\Mes\ne
0$. In order to decouple the unphysical parameter
$\Mes$ from the physical observables, {\it i.e.} $M_h$  in our case,
we modify the top squark
masses by a finite counterterm.  This 
renormalization scheme is  equivalent to the \drbar{}$^\prime$
scheme~\cite{Jack:1994rk,Martin:2001vx}. However,
in the original version of    the \drbar{}$^\prime$ scheme the \epscalar{}
mass is renormalized minimally, whereas we choose the
on-shell scheme. Thus, the finite counterterms we found  differ starting
from two-loops from the ones
given in Ref.~\cite{Martin:2001vx} by
\begin{eqnarray}
  \delta m_{\tilde t_{i}}^2 
  &=& 
  \Mes^2\left(\as\right)^2\bigg[-\frac{\cf^2}{4} +
  \ca\cf\frac{-3 - 4\lMes + \lMgl}{8} 
  \nonumber\\&&\mbox{}
  + \cf\Nq\TF\frac{1 + \lMes}{4}+ \cf\Nt\TF\frac{-1 + \lMt}{4} 
  \bigg]\,,\quad i=1,2\,,
\end{eqnarray}
with $L_{\mu x}=\ln(\mu^2/m_x^2)$ and $\lMes = \ln(\mu^2/\Mes^2)$.
In the second approach we set the   on-shell \epscalar{} mass
to zero. Of course, no finite counterterm is needed in this
case. The results for $M_h$ obtained employing the two methods are
identical, however, the second approach is computationally less involved.
Thus for the practical calculation we set the on-shell 
 \epscalar{} mass to zero.\\
 Let us mention that strictly speaking our renormalization coincides neither 
\drbar{} nor \drbar{}$^\prime$ due to the different treatment of
 the \epscalar{} mass. Nevertheless we use the nomenclature ``\drbar{} scheme''.

Finally, for the mixing angle we have
{\allowdisplaybreaks
\begin{align}
(&\dms)\delta\theta_t^{(1)}=\cf \Cmt
\left(\Mgl\Mt - \frac{\Smt(\Mstu^2 - \Mstd^2)}{4}\right)\frac{1}{\ep}\,,
\nonumber\\
(&\dms)\delta\theta_t^{(2)}=\bigg\{
\cf^2\Cmt\bigg[(\Smt^2 - \Cmt^2)\left(\frac{\Mgl\Mt}{2} -
  \frac{\Smt(\dms)}{16}\right) - \frac{2\Smt\Mgl^2\Mt^2}{\dms} 
\bigg]\nonumber\\
&+
\cf\ca\Cmt\bigg[\frac{-3\Mgl\Mt}{4} + \frac{3\Smt(\dms)}{32}
\bigg]
+
\cf\Nf\TF\Cmt\bigg[\frac{\Mgl\Mt}{2} - \frac{\Smt(\dms)}{16}
\bigg]
\bigg\}\frac{1}{\ep^2}\nonumber\\
&+\bigg\{
\cf^2\Cmt\bigg[-\frac{\Mgl\Mt}{2} + \frac{\Smt(\dms)}{16}\bigg]+
\cf\ca\Cmt\bigg[\frac{3\Mgl\Mt}{4} - \frac{3\Smt(\dms)}{32}
\bigg]\nonumber\\
&+
\cf\Nf\TF\Cmt\bigg[-\frac{\Mgl\Mt}{2} + \frac{\Smt(\dms)}{16}
\bigg]
\bigg\}\frac{1}{\ep}\,.
\label{eq::mixa2l}
\end{align}
}

The two-loop counterterms given in Eqs.~(\ref{eq::mst1}) and (\ref{eq::mixa2l})
can also be derived from the more general results available in the literature
~\cite{Jack:1994kd,Yamada:1994id,Martin:1993zk }.

The  on-shell renormalization constant of the \epscalar{} mass 
to one-loop order is given by
\begin{eqnarray}
(\Mes^2)^{(0)}=\Mes^2 Z_{\Mes^2}\,, \quad Z_{\Mes^2}=1+\as \delta
  Z^{(1)}_{\Mes^2} + {\cal O}(\alpha_s^2)\,,
\end{eqnarray}
where
\begin{eqnarray}
-\Mes^2\delta Z^{(1)}_{\Mes^2}&=&\bigg[
\frac{\ca}{4} (3\Mes^2 + 2\Mgl^2) -\frac{\Nq\TF}{2} (\Mes^2 + 2\msq^2) -
\frac{\Nt\TF}{2}  (\Mes^2 + \Mstu^2 + \Mstd^2 - 2\Mt^2)
\bigg]\frac{1}{\ep}\nonumber\\
&+& \frac{\ca}{4}[(6 + 4\lMes - \lMgl)\Mes^2 + 2 (1 + \lMgl)\Mgl^2]
   \nonumber\\
&-&\frac{\Nq\TF}{2}[ (2 + \lMes)\Mes^2 + 2 (1 + \lMsq)\msquark^2]
\nonumber\\
&-& \frac{\Nt\TF}{2}[
\lMt\Mes^2 + (1 + \lMstu)\Mstu^2 + (1 + \lMstd)\Mstd^2 - 2 (1 + \lMt)\Mt^2]
\,.
\end{eqnarray}

The two-loop  renormalization constant for the top quark mass in the
\drbar{} scheme  is known for long time~\cite{Jones:1983vk,Parkes:1985hj}.
For the convenience of the reader we quote the results which are given by
\begin{eqnarray}
m_t^{(0)}&=&m_t Z_{m_t}\,,\quad \mbox{with}\quad Z_{m_t}=
1+\frac{\alpha_s}{\pi}  \delta
Z_{m_t}^{(1)}+\left(\as\right)^2\delta Z_{m_t}^{(2)}+{\cal O}(\alpha_s^3)
\,,
\end{eqnarray}
and
\begin{eqnarray}
  \delta Z_{m_t}^{(1)} &=&-\cf\frac{1}{2 \ep}\,,\nonumber\\
  \delta Z_{m_t}^{(2)} &=&\left(\frac{1}{8}\cf^2+\frac{3}{16}\ca\cf-\frac{1}{8}
\cf\Nf\TF\right)\frac{1}{\ep^2}\nonumber\\
&+&
 \left(\frac{1}{8}\cf^2-\frac{3}{16}\ca\cf+
\frac{1}{8}\cf\Nf\TF \right)\frac{1}{\ep}\,.
\end{eqnarray}
The same is true for the one-loop gluino mass
 counterterm~\cite{Jones:1983vk,Parkes:1985hj} defined
through
\begin{eqnarray}
m_{\tilde{g}}^{(0)}=m_{\tilde{g}} Z_{m_{\tilde{g}}}\,, \quad 
\mbox{with}\quad Z_{m_{\tilde{g}}}=1+\as\delta Z_{\Mgl}^{(1)} + {\cal
  O}(\alpha_s^2)\,,
\end{eqnarray}
with
\begin{eqnarray}
\delta Z_{\Mgl}^{(1)} &=& \left(-\frac{3}{4}\ca+\frac{1}{2}\Nf\TF
\right)\frac{1}{\ep}\,.
\end{eqnarray}

\section*{\label{sec::appDRmod}Appendix B: Modification of the \drbar{} scheme:
\drbarmod{}}

As discussed in Section~\ref{sec::3loops}, the renormalization constants
 of the top squarks have mass dimension two. Thus, for some hierarchies
 considered in this paper parametrically (and numerically) large 
corrections might appear which are absent in the on-shell scheme. In order to
 cure this problem we introduce additional finite corrections in the
 corresponding renormalization constants which ensure
 that the renormalized result for the Higgs boson mass is free
 of these potentially dangerous contributions.

In the following we provide  analytic expressions for the finite shifts
introduced in the top squark mass counterterms as compared
 to the \drbar{} scheme. According to the discussion in
 Section~\ref{sec::3loops}, one can distinguish three cases
 for the mass hierarchies.

Case (i): $\quad\msq \gg \Msti$, $(i=1,2)$
\begin{eqnarray}
  \left(\frac{\Msti^{\overline{\rm MDR}}}{\Msti}\right)^2 &=& 
  1 - \left(\as\right)^2\cf\Nq\TF
  \frac{\msq^2}{\Msti^2}
  \left(-\frac{1}{2} +\lMsq + \zeta(2) \right)
  \,.
  \label{eq:h4shift}
\end{eqnarray}
The label $\Nq=5$ has been introduced for convenience and for the
logarithms the abbreviation $\lMsq=\ln(\mu^2/\msq^2)$ has been introduced.

Case (ii): $\quad\Mstd \gg \Mstu $
\begin{eqnarray}
  \left(\frac{\Mstu^{\overline{\rm MDR}}}{\Mstu}\right)^2 &=& 
  1 - \left(\as\right)^2\cf\TF
  \frac{\Mstd^2}{\Mstu^2}
  \left(-\frac{1}{4} + \frac{1}{2}\lMstd + \frac{1}{2}\zeta(2) \right)
  \,.
\label{eq:h3shift}
\end{eqnarray}
In this equation we have $\lMstd=\ln(\mu^2/\Mstd^2)$.

Case (iii): $\quad\Mgl \gg \Msti$, $(i=1,2)$ and $\msq\gg \Mgl$
\begin{eqnarray}
  \left(\frac{\Msti^{\overline{\rm MDR}}}{\Msti}\right)^2 &=& 
  1+
  \as\cf\left[1 + \lMgl\right]\frac{\Mgl^2}{\Msti^2}
  +
  \left(\as\right)^2\bigg\{
  \cf^2\left[-\frac{11}{4} - \frac{3}{2}\lMgl 
    + \zeta(2)\right]\frac{\Mgl^2}{\Msti^2}
  \nonumber\\
  &+&
  \ca\cf\left[\frac{21}{8} + \frac{7}{2}\lMgl 
    + \frac{9}{8}\lMgl^2 - \frac{1}{4}\zeta(2)\right]\frac{\Mgl^2}{\Msti^2}
  \nonumber\\
  &+&
  \cf\Nt\TF\left[-\left(2 + 2\lMgl
      + \frac{3}{4}\lMgl^2\right)\frac{\Mgl^2}{\Msti^2} 
    + \left(1-2\zeta(2)\right)\frac{\Mgl(\Mgl-\Mstd)}{\Msti^2}\right]
  \nonumber\\        
  &+&\cf\Nq\TF\bigg[\left(-\frac{5}{8} - \frac{3}{4}\lMgl 
    - \frac{5}{4}\lMsq - \frac{3}{2}\lMgl\lMsq + \frac{3}{4}\lMsq^2 
    + \frac{3}{2}\zeta(2)\right)\frac{\Mgl^2}{\Msti^2}
  \nonumber\\
  &+&
  \left(-\frac{43}{36} - \frac{5}{6}\lMsqgl \right) \frac{\Mgl^4}{\msq^2\Msti^2} 
  +
  \left(-\frac{67}{288} - \frac{7}{24}\lMsqgl \right)\frac{\Mgl^6}{\msq^4\Msti^2} 
  \bigg] 
  \bigg\}
  \,.
  \label{eq:h6shift}
\end{eqnarray}
Here $N_t=1$, $\lMgl=\ln(\mu^2/\Mgl^2)$ and $\lMsqgl=\ln(\msq^2/\Mgl^2)$.

Case (iv): $\quad\Mgl \gg \Msti$, $(i=1,2)$ and $\msq\approx \Mgl$
\begin{eqnarray}
  \left(\frac{\Msti^{\overline{\rm MDR}}}{\Msti}\right)^2 &=&
  1+ \as\cf\left[1 + \lMgl\right]\frac{\Mgl^2}{\Msti^2}
  +
  \left(\as\right)^2\bigg\{\cf^2\left[-\frac{11}{4} - \frac{3}{2}\lMgl +
    \zeta(2)\right]\frac{\Mgl^2}{\Msti^2}\nonumber\\
  & +& \ca\cf \left[\frac{21}{8} + \frac{7}{2}\lMgl + \frac{9}{8}\lMgl^2
    - \frac{1}{4}\zeta(2)\right]\frac{\Mgl^2}{\Msti^2} \nonumber\\
  &+&
  \cf\Nt\TF\left[-\left(2 + 2\lMgl + \frac{3}{4}\lMgl^2\right)
    \frac{\Mgl^2}{\Msti^2} +\left(1 -
      2\zeta(2)\right)\frac{\Mgl(\Mgl-\Mstd)}{\Msti^2}
  \right]
  \nonumber\\
  & +&\cf\Nq\TF\bigg[\left(-\frac{3}{4}\lMgl - \frac{5}{4}\lMsq
    - \frac{3}{2}\lMgl\lMsq + \frac{3}{4}\lMsq^2 +
    \frac{3}{2}\zeta(2)\right)\frac{\Mgl^2}{\Msti^2}
  \nonumber\\
  &-& 4 \zeta(2)\frac{\Mgl(\Mgl-\Mstd)}{\Msti^2}
  -\frac{9}{4}\frac{\msq^2}{\Msti^2}
  \bigg]\bigg\}\,.
  \label{eq:h6bshift}
\end{eqnarray}

All the masses on the r.h.s. of the
Eqs.~(\ref{eq:h4shift}), (\ref{eq:h3shift}), (\ref{eq:h6shift}) and
(\ref{eq:h6bshift}) are \drbar{} 
masses. Cases~(i) and~(ii) are applied for all hierarchies whereas cases~(iii)
and (iv) are only used for (h6) and (h6b), respectively.
Let us also mention that the above formulae are valid for the case  $\Mes=0$.

\section*{Appendix C: SPS scenarios}

In the following table we list the input values for the {\sc msugra}
SPS scenarios as defined in
Refs.~\cite{Allanach:2002nj,AguilarSaavedra:2005pw}. 
All masses are given in GeV and $\textnormal{sign}(\muSUSY)=1$.

\begin{center}
\begin{tabular}{c||c|c|c|c||c|c}
&\multicolumn{4}{|c||}{Points}&\multicolumn{2}{|c}{Slopes}\\
\hline
\multicolumn{1}{c||}{}& $m_0$ & $m_{1/2}$ & $A_0$ & $\tan{\beta}$& $m_0$ & $A_0$\\
\hline
\multicolumn{1}{c||}{SPS1a${}^\prime$} & 70    & 250  & $-$300  &  10  & $-$ &  $-$ \\
\hline
\multicolumn{1}{c||}{SPS1a}  & 100   & 250  &$-$100 & 10 & $0.4m_{1/2}$ &  $-0.4m_{1/2}$ \\
\hline
\multicolumn{1}{c||}{SPS1b}  & 200   & 400  &    0  &  30  & $-$ &  $-$ \\
\hline
\multicolumn{1}{c||}{SPS2}   & 1450  & 300  &    0  &  10  & $2m_{1/2}+850$ & 0 \\
\hline
\multicolumn{1}{c||}{SPS3}   & 90    & 400  &    0  &  10  & $0.25m_{1/2}-10$ & 0 \\
\hline
\multicolumn{1}{c||}{SPS4}   & 400   & 300  &    0  &  50  & $-$ &  $-$ \\
\hline
\multicolumn{1}{c||}{SPS5}   & 150   & 300  & $-$1000 &  5   & $-$ &  $-$ \\
\end{tabular}
\end{center}

\end{document}